\documentclass[
    twocolumn,
    reprint,
    pra,
    noeprint,
    nofootinbib,
]{revtex4-2}

\AtBeginEnvironment{pmatrix}{\everymath{\displaystyle}}
\AtBeginEnvironment{bmatrix}{\everymath{\displaystyle}}

\usepackage{natbib}
\usepackage{microtype}
\usepackage{amsfonts}
\usepackage{mathtools,amssymb}
\usepackage[mathcal]{euscript}
\usepackage[bb=boondox]{mathalpha}

\newcommand{\symbf}{\mathbf}
\newcommand{\symsf}{\mathsf}
\newcommand{\symbfsf}{\mathsf}

\newcommand{\eps}{\varepsilon}
\renewcommand{\phi}{\varphi}

\usepackage[inline]{enumitem}
\usepackage{tabularray}

\usepackage[
    print-unity-mantissa=false,
]{siunitx}
\DeclareSIUnit{\parsec}{pc}
\DeclareSIUnit{\year}{yr}
\DeclareSIUnit{\torr}{Torr}
\DeclareSIUnit{\foot}{ft}
\UseTblrLibrary{siunitx}

\usepackage[
  usenames,
  dvipsnames,
  svgnames,
  x11names,
]{xcolor}
\usepackage[
    colorlinks,
    citecolor=Blue4,
    linkcolor=Blue4,
    urlcolor=Green4!40!black,
    pdfusetitle,
]{hyperref}
\usepackage[capitalise]{cleveref}
\crefname{section}{\S}{\S\S}
\usepackage{verbatim}

\usepackage{graphicx}
\graphicspath{{figures/,../../figures/}}
\usepackage{tikz}
\usetikzlibrary{shapes,arrows,arrows.meta,positioning}
\usepackage{pgfplots}
\usepgfplotslibrary{groupplots,dateplot}
\usetikzlibrary{patterns,shapes.arrows}
\pgfplotsset{compat=newest}

\usepackage{tcolorbox}

\DeclareMathOperator{\real}{Re}
\DeclareMathOperator{\imag}{Im}

\DeclareMathOperator{\sinc}{sinc}

\newcommand{\rmd}{\mathrm{d}}
\newcommand{\rme}{\mathrm{e}}
\newcommand{\rmi}{\mathrm{i}}

\graphicspath{{figures/}}

\setlist{nosep}

\newcommand{\xtal}[1]{{#1}_{\text{X}}}
\newcommand{\pump}[1]{{#1}_{b}}
\newcommand{\fund}[1]{{#1}_{a}}

\newcommand{\?}{\hspace{0.2ex}}

\begin{document}

\title{Precision dynamics of resonantly enhanced optical parametric amplifiers}
\author{Evan D. Hall}
\affiliation{LIGO Laboratory, Department of Physics, Massachusetts Institute of Technology, Cambridge, Massachusetts 02139, USA}

\begin{abstract}
    Optical parametric amplification is a crucial technology for the production of continuous-wave squeezed vacuum, which is now applied to gravitational-wave interferometry and other highly sensitive measurements of optical phase.
    The amplifiers employed in gravitational-wave detection are resonantly enhanced with optical cavities, which introduces nontrivial dynamics to their operation and requires a frequency-domain model of the relations between optical fields.
    In this work, input--output relations between optical fields entering, circulating in, and exiting a model amplifier are solved directly to produce frequency-domain relations that fully incorporate the cavity dynamics.
    We also provide low-order zero--pole--gain expansions for these relations to facilitate their analysis from a feedback control perspective.
    In the limit of small cavity decay rates, our relations are shown to reduce to relations previously derived under a Hamiltonian formalism.
    Expressions for amplifier figures of merit and for squeezed quadrature variances are also analyzed.
    Along the way, we examine how information about the amplifier performance can be extracted from the frequency dependence of the input--output relations, providing characterization methods that are complementary to methods that examine the static amplifier behavior.
\end{abstract}

\maketitle

\section{Introduction}

Continuous-wave squeezed vacuum is now a well-established tool for enhancing sensitivity of many experiments that measure small optical phase shifts at the quantum limit~\cite{Xiao:1987zz,Grangier:1987zz,Mckenzie:2002ir,Vahlbruch:2005nzf,Goda:2008ie,Eberle:2010zz,Mehmet:2010,Grote:2013oio,LIGOScientific:2013pcc,Tse:2019wcy,Virgo:2019juy,Zuo:2020atn,Lough:2020xft,LIGOO4Detector:2023wmz}.
In gravitational wave detection, the preferred method of generating this squeezed vacuum is optical parametric amplification, achieved via interaction of a pump and fundamental fields in a medium with a quadratic nonlinearity.
The medium is integrated into an optical cavity, which enhances the nonlinear interaction and produces squeezing in a well-defined transverse spatial mode.
(See Refs.~\cite{Schnabel:2016gdi,Barsotti:2018hvm,Dwyer:2022vbh} for reviews.)%

The production of squeezed vacuum is often modeled with a Hamiltonian formalism to derive a set of quantum Langevin equations for the time evolution of operators corresponding to the signal and idler modes, coupled to bath modes at rates set by the cavity transmissivities and losses;
input--output relations (transfer functions) are found by transforming into the frequency domain, and field variances characterizing the level of squeezing are finally found by taking appropriate correlation functions of the operators~\cite{Collett:1984ulf,Gardiner:1985uof}.
This same kind of Hamiltonian formalism can be used to model passive optical cavities, which is common in cavity optomechanics~\cite{Aspelmeyer:2013lha}, and is influenced by earlier work on quantization of electronic circuits (e.g., Ref.~\cite{Yurke:1984jot}; see also the illuminating discussion in Ref.~\cite{2003ApOpt..42.4989Z}).
The quantum Langevin equations in this context are typically first order in time, and in the frequency domain therefore yield input--output relations that are low-order rational functions, with coefficients set by the cavity gain and decay rates.

Another approach to cavity dynamics starts from time-domain field relations directly, without recourse to a Hamiltonian, tracking the interaction and propagation of modes as they enter, circulate in, and exit the cavity.
This procedure is typically used for passive optical cavities in gravitational-wave interferometry, and it fully captures features that arise from propagation delays, such as the periodic optical resonance of light at the cavity free spectral range~\cite{2002PhLA..305..239R,Izumi:2016joj}.
Some aspects of the behavior of cavity-enhanced optical parametric amplifiers (OPAs) have been derived under this formalism~\cite{Ganapathy:2022hgu}, including the analysis of force-sensitive cavities with nonlinear enhancement~\cite{Korobko:2017llv,Korobko:2023phs}.
Here we systematically examine the input--output relations of a resonantly enhanced optical parametric amplifier using this procedure, and examine the approximations that put it in correspondence with the results of a Hamiltonian analysis.
There are several reasons to do this:
\begin{enumerate}
    \item The cavity dynamics encodes information about the amplifier (e.g., its coupling and decay rates) in a way that is complementary to the cavity statics, like transmitted and reflected power, and therefore may be used to characterize the cavity in ways that are not subject to the same measurement systematics like mismatch of polarization, coupling into transverse spatial modes, or certain excess losses.
      A full understanding of the dynamics is a prerequisite to such characterization efforts.
    \item The Hamiltonian formalism typically does not include dynamics arising from the finite free spectral range of the cavity.
      Although in gravitational-wave interferometry there is a usually a wide separation of frequency scales between the free spectral range of the squeezed-light-generating OPA (of order gigahertz) and the gravitational-wave observation band (typically below ten kilohertz), various rf control schemes for the OPA operate in the megahertz regime.
      Besides the strict application to audio-band gravitational-wave interferometry, experimental applications may require knowledge of an OPA's frequency response over a wide bandwidth compared to its free spectral range.
    \item Optical parametric amplifiers are often operated in the regime of moderately large decay rates\,---\,perhaps \qtyrange{5}{10}{\%} of the cavity free spectral range\,---\,to facilitate a high degree of squeezed vacuum.
        One therefore might expect the leading-order rate expansions to the cavity dynamics to have corrections at the same fractional order, which could be relevant as the demands on optical parametric amplifier performance and characterization become more exacting.
    \item An analytical model provides a baseline against which numerical computations can be benchmarked, which has been indispensable for precise work with passive cavities~\cite{Izumi:2016joj}, and is a prerequisite to validate more complicated numerical studies that are not analytically tractable (e.g., amplifiers with various nonidealities).
      General-purpose interferometric modeling tools such as \texttt{finesse}~\cite{Finesse} are formulated directly as relations between field modes, rather than Hamiltonians.
\end{enumerate}

Our work is organized as follows.
We first review the form of the parametric interaction in \cref{sec:nonlinear}.
Then, in \cref{sec:fields}, we write the relations between the optical sidebands at various ports of the cavity in frequency domain.
We define thresholds and rates in \cref{sec:rates} and discuss expansion and approximation strategies.
In \cref{sec:io} we derive the full input--output relations between the field modes and also produce low-order zero--pole--gain expressions using Padé expansion.
We show that when the coupling and decay rates are small relative to the cavity free spectral range, these relations nearly reduce to the relations found under the Hamiltonian formalism; the only difference being that our relations include some additional corrections due to the dynamics of the cavity free spectral range.
In \cref{sec:gains} we examine expressions for some figures of merit of the cavity, including longstanding dc measures such as the nonlinear gain and ac responses that can also yield information about the amplifier.
In \cref{sec:noise} we use the input--output relations to examine the frequency-dependent variance of the squeezed vacuum, and show that it reduces to previous expressions found in the literature.
Finally in \cref{sec:conclusion} we point to future work.

\section{Single-pass parametric amplification}
\label{sec:nonlinear}

To start, we derive the interaction of optical signal and idler modes inside the $\chi^{(2)}$-nonlinear element of the amplifier.
Our goal in this section is to derive the parametric interaction of signal and idler modes in a plane-wave geometry, compare it to expressions that have been presented previously, and point out some relevant assumptions that underlie the interaction model in the context of free-space cavity amplifiers.
These results are needed to proceed with a full model of a cavity parametric amplifier in subsequent sections.

When pumped with a coherent optical field at angular frequency $\pump{\omega}$, the amplifier's nonlinear medium couples the signal and idler modes in the vicinity of the fundamental field, which has angular frequency $\fund{\omega} = \pump{\omega}/2$.
We denote by $a_+$ the complex amplitude of the optical mode at $\fund{\omega}+\Omega$ (the upper sideband, or the signal), and by $a_-$ the complex amplitude of the mode at $\fund{\omega}-\Omega$ (the lower sideband, or the idler) with the assumption $|\Omega| \ll \fund{\omega}$.
For broadband squeezed light enhancement of optical interferometers, this assumption is usually well satisfied: the fundamental field typically has frequency $\fund{\omega}/2\pi > \qty{100}{\THz}$, while the sideband frequencies of interest for most gravitational-wave searches are $|\Omega/2\pi| < \qty{10}{\kHz}$; the assumption is still well satisfied even for searches extending to megahertz or gigahertz.
The quantum-mechanical theory of the parametric interaction fits within the well-known two-photon formalism \cite{Caves:1985zz,Schumaker:1985zz} (see also Ref.~\cite{Danilishin:2012fa} for a review).

\subsection{Structure of the parametric interaction matrix}
\label{sec:single pass derivation}

The classical analysis of optical parametric amplification in quadratically nonlinear media has a standard treatment given by, e.g., \textcite[Ch.~9]{YarivYehPhotonics} or \textcite[Ch.~2]{BoydNonlinear}.
We suppose the pump mode and the fundamental modes are collinear plane waves propagating in the $+z$ direction.
In a linear medium, the mode amplitudes evolve independently, but in a $\chi^{(2)}$-nonlinear medium, the induced polarization density depends on products of the modes' electric field amplitudes, which couples their evolution as they propagate through the medium.
We henceforth define the medium as a crystal with a longitudinal extent from $0$ to $+\xtal{l}$; its refractive index is $\fund{n}$ for the fundamental modes and $\pump{n}$ for the pump mode.
The crystalline material (including any periodic poling), choice of wavelength, and the arrangement of the modes relative to the crystal axes define an effective second-order nonlinearity $d_{\text{eff}}$, with dimensions of inverse electric field.

If the power depleted from the pump mode by the nonlinear interaction is negligible, then
the resulting coupling is solely between the complex amplitudes of the upper and lower sideband modes.
Collecting the complex sideband amplitudes $a_+$ and $a_-^\dagger$ into a column vector $\symbf{a}$, the evolution of the amplitudes as a function of displacement $z$ for a plane-wave geometry is given by~\cite{YarivYehPhotonics}
\begin{equation}
  \frac{\rmd \symbf{a}(z)}{\rmd{z}}
        = \begin{bmatrix} 0 & -\frac{\rmi\gamma}{2} \rme^{-\rmi\Delta k\,z}\\ +\frac{\rmi\gamma^*}{2} \rme^{+\rmi\Delta k\,z} & 0 \\ \end{bmatrix}
        \symbf{a}(z)
   \label{eq:opa diff eq}
\end{equation}
with
\begin{equation}
    \Delta k = \pump{k} - 2\fund{k} - k_0
    \label{eq:Delta k}
\end{equation}
being the mismatch between the fundamental wavenumber $\fund{k} = \fund{\omega} \fund{n} /c$ and pump wavenumber $\pump{k} = \pump{\omega} \pump{n} / c$ within the crystal; $k_0$ accounts for effects such as periodic poling~\cite{1995JOSAB..12.2102M}.
The nonlinear interaction per unit length is given by
\begin{equation}
    \rmi\gamma = \frac{\rmi\fund{\omega}}{c\fund{n}} d_{\text{eff}} \pump{E} \equiv - |\gamma| \rme^{-2\rmi\pump{\psi}}
    \label{eq:g},
\end{equation}
where $\pump{E}$ is the complex amplitude of the electric field of the pump\,---\,having phase $-2\pump{\psi}$ and magnitude related to the intensity by $\pump{I} = \tfrac{1}{2}\eps_0 \pump{n} c |\pump{E}|^2$\,---\,and $c$ is the vacuum speed of light.

\begin{widetext}
\cref{eq:opa diff eq} has a standard solution found by making a change of variable and then supposing an ansatz of complex exponentials~\cite{ByerOPO,YarivYehPhotonics,BoydNonlinear}.
A slightly different route to the solution is given in \cref{sec:su11}, where the Lie algebra structure of the nonlinear interaction is emphasized.
In this route it becomes clear that the solution must belong to $\mathrm{SU}(1,1)$, which places restrictions on its functional form.
In either case, the solution to \cref{eq:opa diff eq} that takes a pair of sidebands $\symbf{a}(0)$ at the crystal input and transforms them into a pair of sidebands $\symbf{a}(+\xtal{l})$ at the crystal output is the matrix product
\begin{equation}
    \xtal{\symsf{H}} = %
    \begin{bmatrix}
      \exp\left(-\rmi\frac{\Delta k\,\xtal{l}}{2}\right) & 0 \\
      0 & \exp\left(+\rmi\frac{\Delta k\,\xtal{l}}{2}\right)
    \end{bmatrix}
    \begin{bmatrix}
        \left(\cosh\frac{\mu\xtal{l}}{2} + \frac{\rmi\Delta k}{\mu} \sinh\frac{\mu\xtal{l}}{2}\right) &
        -\frac{\rmi\gamma}{\mu} \sinh\frac{\mu\xtal{l}}{2} \\
        +\frac{\rmi\gamma^*}{\mu} \sinh\frac{\mu\xtal{l}}{2} &
        \left(\cosh\frac{\mu\xtal{l}}{2} - \frac{\rmi\Delta k}{\mu} \sinh\frac{\mu\xtal{l}}{2}\right) \\
    \end{bmatrix}
    \label{eq:HX}
\end{equation}
with $\mu = \sqrt{|\gamma|^2 - (\Delta k)^2}$.%
\footnote{For the particular case $|\gamma| = |\Delta k|$, \cref{eq:HX} can be evaluated by taking the right-sided limit $\mu \rightarrow 0^+$.}
The $\mathrm{SU}(1,1)$ structure of $\xtal{\symsf{H}}$ is evinced by the fact that both of its component matrices have unit determinant, and the diagonal entries in each matrix form a complex conjugate pair, as do the off-diagonal entries in each matrix.

In some circumstances it is convenient to work with an alternate functional form of the parametric interaction matrix:
\begin{equation}
    \xtal{\symsf{H}} = 
    \begin{bmatrix}
      \exp\left(-\rmi\frac{\Delta k\,\xtal{l}}{2}\right) & 0 \\
      0 & \exp\left(+\rmi\frac{\Delta k\,\xtal{l}}{2}\right)
    \end{bmatrix}\begin{bmatrix}
        \rme^{-\rmi\zeta}\cosh\Gamma & -\frac{\rmi \gamma}{|\gamma|} \sinh\Gamma \\
        +\frac{\rmi \gamma^*}{|\gamma|} \sinh\Gamma & \rme^{+\rmi\zeta}\cosh\Gamma
    \end{bmatrix},
    \label{eq:HX hyperbolic}
\end{equation}
where $\Gamma$ and $\zeta$ are defined via the implicit relations
\begin{equation}
    \sinh\Gamma = \frac{|\gamma|\xtal{l}}{2} \frac{\sinh(\mu\xtal{l}/2)}{\mu\xtal{l}/2}
    \label{eq:Gamma}
\end{equation}
and
\begin{equation}
  -\tan\zeta = \frac{\Delta k\,\xtal{l}}{2}\frac{\tanh(\mu\xtal{l}/2)}{\mu\xtal{l}/2}.
  \label{eq:zeta}
\end{equation}
\end{widetext}
Although $\mu$ may be real or imaginary, $\Gamma$ is always nonnegative and $\zeta$ is always real.
For either form of $\xtal{\symsf{H}}$, when $\Delta k \rightarrow 0$, we recover the result (using \cref{eq:g})
\begin{equation}
    \xtal{\symsf{H}} \xrightarrow{\Delta k \rightarrow 0}
    \begin{bmatrix}
        \cosh{\Gamma} & \rme^{-2\rmi\pump{\psi}} \sinh{\Gamma} \\
        \rme^{+2\rmi\pump{\psi}} \sinh{\Gamma} & \cosh{\Gamma}
    \end{bmatrix}
    \label{eq:HX Ganapathy}
\end{equation}
with $\Gamma \xrightarrow{\Delta k \rightarrow 0} |\gamma|\xtal{l}/2$.
This agrees with a number of previous expressions in the gravitational-wave literature~\cite{Caves:1985zz,Ganapathy:2022hgu,Chelkowski:2007teo}.

\subsection{Application to cavity amplifiers}

It is worth pausing to remark on several features of this solution and its applicability to cavity parametric amplifiers.
First, $\xtal{\symsf{H}} \rightarrow \symsf{1}$ when $\gamma \rightarrow 0$, which means that the finite light travel time of the fields traveling through the crystal length $\xtal{l}$ has not been accounted for.
One can approximately include this effect by separately applying a matrix of time delays to the final sideband vector $\symbf{a}(+\xtal{l})$.
This amounts to the assumption that the nonlinear interaction occurs on a timescale much faster than the other timescales in the amplifier, such as the travel time around the cavity.
Indeed we will proceed with this assumption in the next sections, although we note it will not be true in devices where the path through the nonlinear element forms a significant fraction of the total round-trip path in the cavity.

Second, the interaction here has assumed the modes are plane waves.
In reality, they are paraxial beams with a Gaussian transverse spatial profile, which introduces additional effects to the nonlinear interaction, especially when (as is usually the case) the beams are strongly focused to enhance the interaction~\cite{1968JAP....39.3597B,2007OExpr..15.7211L,BoydNonlinear}.
In particular, the interaction per unit length $\gamma$ becomes dependent on the axial coordinate $z$, and the modes acquire additional axially dependent phase shifts.
We will not consider these effects here, and the particular form of $\xtal{\symsf{H}}$ may need to be found numerically in general.
However, we argue in \cref{sec:su11} that for a wide class of interaction effects, $\xtal{\symsf{H}}$ still must belong to $\mathrm{SU}(1,1)$, and thus no matter how complicated the mode geometry, one needs \emph{only} three real numbers to characterize the overall parametric interaction: a nonnegative hyperbolic parameter and two real angles.
In the plane wave expression \cref{eq:HX hyperbolic} the hyperbolic parameter is $\Gamma$, and the two real angles are $\Delta k \xtal{l}/2 + \zeta$ and $\Delta k \xtal{l}/2 + 2\pump{\psi}$.

Third, in addition to questions of focusing effects in Gaussian beams, in some applications it is desirable to consider parametric amplification in transverse spatial modes of the optical field beyond the Gaussian $\text{TEM}_{00}$ mode (e.g., Ref.~\cite{Heinze:2022vky}), and in this case the generalization of the above analysis is formally straightforward: if $p$ spatial modes are considered, the sideband vectors are enlarged from dimension $2$ to dimension $2p$, and the interaction matrix $\xtal{\symsf{H}}$ is enlarged from $2\times2$ to $2p\times2p$.
The exact form of the interaction matrix, and the quantities needed to parametrize it, depend on the nature of the intermode coupling and are left for future work.

\subsection{Inclusion of loss}
\label{sec:absorption}

Consideration of internal loss is critical to modeling of a cavity amplifier, because it alters the statics, dynamics, and therefore the performance of the amplifier.
The inclusion of a crystal inside the cavity makes this crystal a likely source of loss.
Crystal loss can be incorporated in the derivation of the parametric interaction by adding $-\mathcal{A}/2$ to the diagonals of \cref{eq:opa diff eq}, where $\mathcal{A}$ is the power loss per unit length.
As described in \cref{sec:su11}, the resulting nonlinear interaction is identical to the lossless case except for the introduction of a prefactor $\rme^{-\mathcal{A}\xtal{l}/2}$ in front of every element of $\xtal{\symsf{H}}$.
To preserve unitarity, the channels mediating the loss\,---\,i.e., sites of scattering or absorption within the crystal\,---\,must also couple fluctuations into the signal and idler fields, and we include this effect by introducing a vacuum mode $\symbf{a}^\text{(X)}$ superposed after the nonlinear interaction, so that
\begin{equation}
  \symbf{a}(+\xtal{l}) = \underbrace{\rme^{-\mathcal{A}\xtal{l}/2}}_{\equiv\xtal{t}} \xtal{\symsf{H}} \symbf{a}(0) - \underbrace{\sqrt{1 - \rme^{-\mathcal{A} \xtal{l}}}}_{\equiv\xtal{r}} \symbf{a}^\text{(X)}.
  \label{eq:aX}
\end{equation}
This construction also implies the coupling of the signal and ilder fields to an outgoing loss field $\symbf{a}^{(\text{X}')}$, although we do not need to keep track of it to describe the amplifier operation and will not consider it further.
The quantities $\xtal{t}$ and $\xtal{r}$ defined in \cref{eq:aX} can easily be modified to include losses from, e.g., the antireflection coating on the outgoing face of the crystal.
Loss at the other antireflection coating of the crystal could be included as a separate interaction before the nonlinear coupling.

\begin{figure*}[t]
    \centering
    \includegraphics[width=5in]{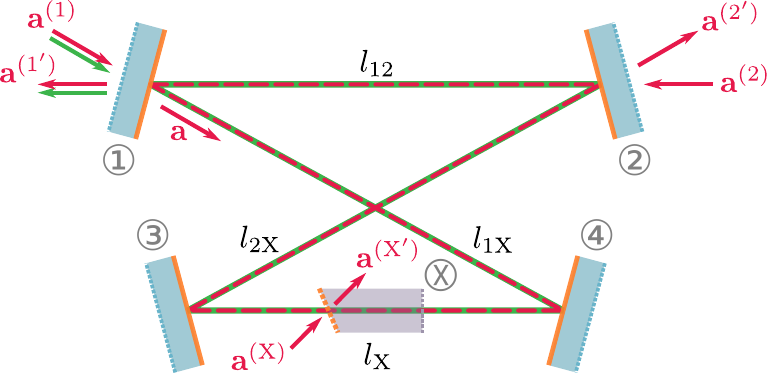}
    \caption{Resonant optical parametric amplifier formed of a quadratically nonlinear crystal (X) and four mirrors, similar to the amplifier used in Advanced LIGO~\cite{2015NatSR...518052W,T1700104,T2000273}.
    Mirrors~1 and~2 are partially transmissive, and mirrors~3 and~4 are assumed to be perfectly reflecting.
    Ingoing and outgoing sideband fields for the fundamental modes are indicated.
    The length $l_{12}$ covers the distance from mirror~1 to mirror~2; $l_{1\text{X}}$ covers the distance from mirror~1 to the flat face of the crystal; $l_{2\text{X}}$ covers the distance from mirror~2 to the wedged face of the crystal; and $l_{\text{X}}$ covers the physical distance through the crystal excluding the wedge.
    We have drawn the optical loss in the crystal as a single beamsplitter operation on the final (wedged) face of the crystal, which is sufficient to accommodate loss due to absorption along the total crystal length.}
    \label{fig:vopo}
\end{figure*}

\section{Field relations for resonant amplifiers}
\label{sec:fields}

We now want to use the single-pass sideband transfer function matrix $\xtal{\symsf{H}}$ found above (\cref{eq:HX}) to derive the transfer functions for an optical parametric amplifier that uses a free-space optical cavity to resonantly enhance fundamental sidebands interacting with the nonlinear crystal.
Rather than working with sideband fields as continuous functions of displacement, we will consider a set of fields defined at some discrete points along the optical beam trajectories, and we work in the Laplace domain with $s = \rmi\Omega$ (refer to \cref{sec:sidebands} for conventions).
At this point we specialize slightly to the type of amplifier adopted for injection of squeezed vacuum in Advanced LIGO~\cite{Chua:2011qha,2015NatSR...518052W} (also applied to \qty{2}{\um} laser wavelength for possible use in future gravitational-wave detectors~\cite{Mansell:2018tbn}).
This is a traveling-wave cavity, in which the fundamental and pump modes circulate in a single direction.
A diagram is shown in \cref{fig:vopo} and parameters are given in \cref{tab:vopo}.
Although we have specialized to a particular amplifier topology, the tools and procedure presented here are broadly applicable, and could be applied with modification to other cavity-enhanced parametric amplifiers, including those based on linear cavities.
Refs.~\cite{Vahlbruch:2010cy,Mehmet:2011je,Mehmet:2018vxw,Darsow-Fromm:2021zif}, for example, describe linear cavity amplifiers employed for gravitational-wave detectors.

The salient aspects of the amplifier here are that one mirror (labeled 1) is chosen to have a transmissivity that dominates the cavity's overall decay rate, so that ingoing sidebands $\symbf{a}^{(1)}$ are highly overcoupled to the cavity;
this facilitates the production of squeezed vacuum in the reflected field $\symbf{a}^{(1')}$.
A second mirror (2) has a small but nonzero transmissivity to enable sidebands to be injected ($\symbf{a}^{(2)}$) and extracted ($\symbf{a}^{(2')}$) for control and diagnostic purposes.
The sidebands internal to the cavity and directed outward from mirror~1 are denoted by $\symbf{a}$.
As discussed in \cref{sec:absorption}, crystal loss is considered as a third port (X) that unavoidably adds unsqueezed vacuum sidebands $\symbf{a}^{(\text{X})}$.
This amplifier is also resonant for the pump field $\symbf{b}$, although we will not track its evolution in our analysis.
However, to enforce dual resonance of the pump and fundamental fields, the crystal has a wedged face to enable dispersive adjustment of the relative phase delay between the pump and fundamental.
Other methods of enforcing dual resonance\,---\,e.g., by engineering temperature gradients~\cite{SchoenbeckThesis,HagemannThesis} in the crystal or applying strain\,---\,could also be accommodated in this formalism.

In \cref{sec:nonlinear} we have already written down the matrix $\xtal{\symsf{H}}$ describing the nonlinear coupling of intracavity sidebands.
To construct a full mathematical model of the amplifier with its cavity, we need to keep track of time (or phase) delays between the injected, circulating, and extracted sidebands, and interactions between the sideband fields at the partially transmissive (beamsplitter) mirrors.
To begin, we account for phase delays due to the crystal wedge and other static round-trip cavity phase offsets, such as mirror reflection phases, via the variable $\fund{\phi}$.
In the Laplace domain, the matrix describing these delays, along with the propagation delay $\xtal{\tau} = \xtal{l}\fund{n}/c$ through the portion of the crystal excluding the wedge, is
\begin{subequations}
  \label{eq:Lprop}
\begin{equation}
    \xtal{\symsf{L}}(s) = \rme^{-s\xtal{\tau}}\begin{bmatrix}
        \rme^{-\rmi(\fund{\omega}\xtal{\tau} + \fund{\phi})} & 0 \\
        0 & \rme^{+\rmi(\fund{\omega}\xtal{\tau} + \fund{\phi})} \\
    \end{bmatrix}.
    \label{eq:LX}
\end{equation}
The Laplace-domain matrices describing the effect of free-space propagation delays from mirror~2 to mirror~1, from mirror~1 to the flat face of the crystal, and from the wedged face of the crystal to mirror~2, are, respectively,
\begin{align}
    \symsf{L}_{12}(s) &= +\rme^{-s\tau_{12}}\begin{bmatrix}
        \rme^{-\rmi \fund{\omega} \tau_{12}} & 0 \\
        0 & \rme^{+\rmi \fund{\omega} \tau_{12}} \\
    \end{bmatrix} \\
    \symsf{L}_{1\text{X}}(s) &= -\rme^{-s\tau_{1\text{X}}}\begin{bmatrix}
        \rme^{-\rmi \fund{\omega} \tau_{1\text{X}}} & 0 \\
        0 & \rme^{+\rmi \fund{\omega} \tau_{1\text{X}}} \\
    \end{bmatrix} \\
    \symsf{L}_{2\text{X}}(s) &= -\rme^{-s\tau_{2\text{X}}}\begin{bmatrix}
        \rme^{-\rmi \fund{\omega} \tau_{2\text{X}}} & 0 \\
        0 & \rme^{+\rmi \fund{\omega} \tau_{2\text{X}}} \\
    \end{bmatrix},
\end{align}
\end{subequations}
where $\tau_{12} = l_{12}/c$, $\tau_{1\text{X}} = l_{1\text{X}}/c$, and $\tau_{2\text{X}} = l_{2\text{X}}/c$ are the propagation times for these paths.
The minus signs in $\symsf{L}_{1\text{X}}$ and $\symsf{L}_{2\text{X}}$ account for the reflections off mirrors~3 and~4, which are assumed to have amplitude reflectivities of $-1$ and are not included as separate propagation steps in this derivation.

The total round-trip detunings of the fundamental and pump fields (not including audio-frequency delays in any sidebands), are, respectively, 
\begin{align}
    \fund{\Phi} &= \frac{\fund{\omega}}{c} \left( l_{12} + l_{1\text{X}} + l_{2\text{X}} + \fund{n}\xtal{l}\right) + \fund{\phi} \\
    \pump{\Phi} &= \frac{\pump{\omega}}{c} \left( l_{12} + l_{1\text{X}} + l_{2\text{X}} + \pump{n}\xtal{l}\right) + \pump{\phi}
\end{align}
Therefore, since $\pump{\omega} = 2\fund{\omega}$,
\begin{align}
    \fund{\Phi} - \frac{\pump{\Phi}}{2} = - \frac{(\pump{k} - 2\fund{k})\, \xtal{l}}{2} + \fund{\phi} - \frac{\pump{\phi}}{2}.
    \label{eq:dPhi}
\end{align}
We will assume the pump is maintained at resonance ($\pump{\Phi} = 0 \pmod{2\pi}$), which is usually achieved by feedback control.
\cref{eq:dPhi} then encodes the operational requirements of a resonant optical parametric amplifier of the type in \cref{fig:vopo}: first, the crystal temperature must be adjusted to set $\pump{k} - 2\fund{k} = k_0$, which maximizes the nonlinear coupling $\Gamma$ according to \cref{eq:Gamma}; then, the transverse position of the crystal relative to the optical modes must be adjusted to set $\fund{\phi} - \pump{\phi}/2 = (k_0 \xtal{l}/2) \pmod{2\pi}$, which brings the fundamental fields to resonance according to \cref{eq:dPhi}.

In addition to descriptions of the nonlinear interaction and the propagation delays, we need to establish the boundary conditions at the dielectric interfaces that passively couple the various modes mentioned above.
At an interface with amplitude reflectivity $r_k$ and transmissivity $t_k$, both positive and satisfying $0 \le r_k^2 + t_k^2 \le 1$, the ingoing sidebands $\symbf{a}^{(i)}$ and outgoing sidebands $\symbf{a}^{(i')}$ on the vacuum side are related to the ingoing sidebands $\symbf{a}^{(j)}$ and outgoing sidebands $\symbf{a}^{(j')}$ on the substrate side by a beamsplitter interaction
\begin{subequations}
\label{eq:beamsplitter}
\begin{align}
    \symbf{a}^{(i')} &= -r_k \symbf{a}^{(i)} + t_k \symbf{a}^{(j)} \\
    \symbf{a}^{(j')} &= +t_k \symbf{a}^{(i)} + r_k \symbf{a}^{(j)};
\end{align}
\end{subequations}
this interaction holds equally well in the time or frequency domains.
It applies to the dielectric interfaces of the four cavity mirrors, and as mentioned in \cref{sec:absorption}, we also apply it to the antireflection coating on the wedged face of the nonlinear crystal to account for the total loss in this crystal.

\begin{widetext}
Now referring to \cref{fig:vopo}, we can write down a system of linear equations in the Laplace domain relating the fundamental sideband vectors to each other using the beamsplitter interaction (\cref{eq:beamsplitter}), the propagation delays (\cref{eq:Lprop}), and the nonlinear interaction (\cref{eq:HX}).
For the intracavity sideband vector, this results in the Laplace-domain relation
\begin{align}
    \symbf{a} &= t_1 \symbf{a}^{(1)} - r_1 t_2 \symsf{L}_{12}\symbf{a}^{(2)} + \symsf{L}_{1\text{X}}^{-1} \? \symsf{G} \? \symsf{L}_{1\text{X}}\symbf{a} - r_1 r_2 \xtal{r} \symsf{L}_{12} \symsf{L}_{2\text{X}} \symbf{a}^{(\text{X})} \label{eq:a0} \\
    &= \symsf{L}_{1\text{X}}^{-1} \left(\symsf{1} - \symsf{G}\right)^{-1} \symsf{L}_{1\text{X}} \left(t_1 \symbf{a}^{(1)} - r_1 t_2 \symsf{L}_{12} \symbf{a}^{(2)} - r_1 r_2 \xtal{r} \symsf{L}_{12} \symsf{L}_{2\text{X}}\symbf{a}^{(\text{X})} \right),
    \label{eq:a}
\end{align}
where we have defined the round-trip gain
\begin{equation}
    \symsf{G}(s) = r_1 r_2 \xtal{t} \symsf{L}_{1\text{X}}(s) \symsf{L}_{12}(s) \symsf{L}_{2\text{X}}(s) \xtal{\symsf{L}}(s) \xtal{\symsf{H}}
    = r_1 r_2 \xtal{t} \rme^{-s\fund{\tau}} \begin{bmatrix}
        \rme^{-\rmi\fund{\Phi} - \rmi\Delta k \xtal{l} / 2 - \rmi\zeta} \cosh{\Gamma} &
        \rme^{-\rmi\fund{\Phi} - \rmi\Delta k \xtal{l} / 2 - 2\rmi\pump{\psi}} \sinh{\Gamma} \\
        \rme^{+\rmi\fund{\Phi} + \rmi\Delta k \xtal{l} / 2 + 2\rmi\pump{\psi}} \sinh{\Gamma} &
        \rme^{+\rmi\fund{\Phi} + \rmi\Delta k \xtal{l} / 2 + \rmi\zeta} \cosh{\Gamma} \\
    \end{bmatrix}
    \label{eq:G}
\end{equation}
and we can define the internal gain $\underline{\symsf{G}}(s) = \left(\symsf{1} - \symsf{G}(s)\right)^{-1}$.
The reflected field on the overcoupled side of the cavity, which contains the squeezed modes, is
\begin{align}
    \symbf{a}^{(1')} &= r_1 \symbf{a}^{(1)} + t_1 \symsf{L}_{12} \left(t_2 \symbf{a}^{(2)} - r_2 \xtal{t} \symsf{L}_{2\text{X}} \xtal{\symsf{L}} \symsf{H}_{\text{X}} \symsf{L}_{1\text{X}} \symbf{a} + r_2 \xtal{r} \symsf{L}_{2\text{X}} \symbf{a}^{(\text{X})} \right) \\
    &= \symsf{L}_{1\text{X}}^{-1} \left(r_1 \symsf{1} - \frac{t_1^2}{r_1} \overline{\symsf{G}} \right) \symsf{L}_{1\text{X}} \symbf{a}^{(1)} + t_1 t_2 \symsf{L}_{1\text{X}}^{-1} \? \underline{\symsf{G}} \? \symsf{L}_{1\text{X}} \symsf{L}_{12}\symbf{a}^{(2)} + t_1 r_2 \xtal{r} \symsf{L}_{1\text{X}}^{-1} \? \underline{\symsf{G}} \? \symsf{L}_{1\text{X}} \symsf{L}_{12} \symsf{L}_{2\text{X}} \symbf{a}^{(\text{X})}
    \label{eq:a1prime}
\end{align}
where we have defined the closed-loop gain $\overline{\symsf{G}}(s) = \symsf{G}(s) \left(\symsf{1} - \symsf{G}(s)\right)^{-1}$.
The propagation delay matrices $\left\{\symsf{L}_j\right\}$ appearing explicitly in \cref{eq:a1prime} and similar equations are typically uninteresting because they merely amount to a time translation of the fields external to the amplifier; on the other hand, the sequence of delays in the round-trip gain \cref{eq:G} is crucial for setting the travel time around the cavity, and hence its free spectral range and all other frequency scales in the input--output relations.

Meanwhile, the reflected field at the other external port, which is typically used for control purposes, is
\begin{align}
    \symbf{a}^{(2')} &= r_2 \symbf{a}^{(2)} + t_2 \symsf{L}_{2\text{X}} \left(-\xtal{r} \symbf{a}^{(\text{X})} + \xtal{t} \xtal{\symsf{L}} \xtal{\symsf{H}} \symsf{L}_{1\text{X}} \symbf{a} \right) \\
    &= \symsf{L}_{12}^{-1} \symsf{L}_{1\text{X}}^{-1} \left(r_2\symsf{1} - \frac{t_2^2}{r_2} \overline{\symsf{G}} \right) \symsf{L}_{1\text{X}} \symsf{L}_{12} \symbf{a}^{(2)} + \frac{t_1 t_2}{r_1 r_2} \symsf{L}_{12}^{-1} \symsf{L}_{1\text{X}}^{-1} \overline{\symsf{G}} \? \symsf{L}_{1\text{X}} \symsf{L}_{12} \symbf{a}^{(1)} - t_2 \xtal{r} \symsf{L}_{12}^{-1} \symsf{L}_{1\text{X}}^{-1} \? \underline{\symsf{G}} \? \symsf{L}_{1\text{X}} \symsf{L}_{12} \symsf{L}_{2\text{X}} \symbf{a}^{(\text{X})}.
    \label{eq:a2prime}
\end{align}
\end{widetext}
Formally, the expressions for the outgoing fields $\symbf{a}^{(1')}$ and $\symbf{a}^{(2')}$ in terms of the respective ingoing fields $\symbf{a}^{(1)}$ and $\symbf{a}^{(2)}$ are quite similar, requiring only an exchange of indices $1 \leftrightarrow 2$ on the reflectivity and transmissivity coefficients and the appearance of some additional propagation delays $\symsf{L}_j$.
But because the fields are overcoupled at mirror~1 and undercoupled at mirror~2, the phenomenology of the two reflections is quite different, which we shall demonstrate later.
Conversely, contributions to the outgoing fields from each ingoing field at the opposite mirror differ formally: $\symbf{a}^{(2)}$ couples to $\symbf{a}^{(1')}$ via a transmission function proportional to the internal gain $\underline{\symsf{G}}$, while $\symbf{a}^{(1)}$ couples to $\symbf{a}^{(2')}$ via a transmission function proportional to the closed-loop gain $\overline{\symsf{G}}$.
However, we shall see that as long as the cavity finesse is reasonably high, the internal gain and the closed-loop gain are not significantly different.

\section{Threshold, rates, and approximation strategy}
\label{sec:rates}

This section establishes expressions for the cavity's threshold and its decay and gain rates.
It also describes the approximation techniques used to reduce the full input--output relations to low-order zero--pole--gain expressions.

\subsection{Threshold}

If a pair of coherent sidebands $\dfrac{1}{\sqrt{2}} \begin{bmatrix} 1 \\ 1 \end{bmatrix}$, which together have unit power, are sent for a single trip around the cavity, they are amplified to $\dfrac{\underline{\symsf{G}}}{\sqrt{2}} \begin{bmatrix} 1 \\ 1 \end{bmatrix}$ and therefore attain a power
\begin{equation}
    \frac{1}{2} \begin{bmatrix} 1 & 1 \end{bmatrix} \symsf{G}^\dagger \symsf{G} \begin{bmatrix} 1 \\ 1 \end{bmatrix}
        = r^2 \rme^{+2\Gamma},
\end{equation}
with $\symsf{G}$ given by \cref{eq:G}, assuming $\Delta k = 0$ and $\fund{\Phi} = 0$.
Here we have defined
\begin{equation}
  r = r_1 r_2 \xtal{t}.
  \label{eq:r}
\end{equation}
The special value of the nonlinear coupling
\begin{equation}
    \Gamma_{\star} = -\ln{r}
    \label{eq:threshold}
\end{equation}
defines the amplifier's threshold, where the power gained in the nonlinear interaction exactly balances the power lost in each trip around the cavity.
\cref{eq:threshold} agrees with previous expressions~\cite{Korobko:2017llv,GanapathyThesis}, and agrees with much of the existing literature derived under the Collett--Gardiner model once expressed in terms of coupling and decay rates (see below).

\subsection{Rates}

In a passive cavity, it is typical to approximate the dynamics by expanding in terms of the power transmissivities and losses.
However, cavity-enhanced parametric amplification depends sensitively on the nonlinear coupling relative to threshold.
Because of this, we need to work directly with the decay rates~\cite{WadeThesis,MansellThesis}
\begin{subequations}
\begin{align}
    \kappa &= -\frac{\ln{r}}{\fund{\tau}} \\
    \kappa_1 &= -\frac{\ln{r_1}}{\fund{\tau}} \\
    \kappa_2 &= -\frac{\ln{r_2}}{\fund{\tau}} \\
    \xtal{\kappa} &= -\frac{\ln{\xtal{t}}}{\fund{\tau}} = \frac{\mathcal{A}\xtal{l}}{2\fund{\tau}}
\end{align}
\label{eq:kappa}
\end{subequations}
and the nonlinear coupling rate
\begin{equation}
    g = \Gamma/\fund{\tau}.
\end{equation}
With these definitions, \cref{eq:r} implies the exact relation $\kappa = \kappa_1 + \kappa_2 + \kappa_{\text{X}}$,
and we can define a normalized coupling
\begin{equation}
    x = \frac{g}{\kappa} = -\frac{\Gamma}{\ln{r}},
\end{equation}
for which threshold occurs at $g_\star / \kappa = 1$.
Later, we will also need the escape efficiency
\begin{equation}
    \eta = \frac{\kappa_1}{\kappa} = \frac{\ln{r}_1}{\ln{r}}.
    \label{eq:ee}
\end{equation}
In some other literature, the escape efficiency is defined in terms of the ratio of the power transmissivity $t_1^2$ to the sum of all the cavity's power transmissivities and losses.
This is approximately true so long as all transmissivities and losses are small, but here we will keep to the exact expression \cref{eq:ee}.

\subsection{Expansions and approximations}
\label{sec:approximations}

In the next section, we will write down the exact frequency-domain relations for the gain, transmission, and reflection matrices for various fields entering, circulating in, and exiting the amplifier.
However, the understanding of the frequency response is enhanced by examining the system's poles and zeros.
An exact pole--zero expansion of the frequency response would yield an infinite number of poles and zeros, because the cavity has an infinite comb of resonant frequencies separated by the free spectral range, corresponding to the angular frequency $\Omega_{\text{FSR}} = 2\pi/\fund{\tau}$~\cite{2002PhLA..305..239R}.
On the other hand, for continuous-wave applications where the desire is to generate broadband squeezed light, we are typically interested in the frequency response for sideband frequencies within one free spectral range of the optical carrier; i.e., the regime $|s\fund{\tau}| < 2\pi$.

As we shall see, the frequency dependence of the gain, transmission, and reflection matrices occurs solely via the expression $r\rme^{-s\fund{\tau}}$, which describes the round-trip reflectivity of the amplifier cavity along with the round-trip phase delay of the optical sidebands relative to the fundamental carrier frequency at $\fund{\omega}$.
We shall expand it via its $[1/1]$ Padé approximant:
\begin{equation}
    r\rme^{-s\fund{\tau}} \xrightarrow{\text{Padé}} r\times\frac{1 - s\fund{\tau}/2}{1 + s\fund{\tau}/2},
    \label{eq:Padé}
\end{equation}
which comprises a left half-plane pole at $s = -2/\fund{\tau}$ and a right half-plane zero at $s = +2/\fund{\tau}$.
By making this expansion, we will arrive at low-order rational function approximations (generally of second order) that can be factorized into zero--pole--gain form.
We note that along $s = \rmi\Omega$, the expansion in \cref{eq:Padé} preserves the unitarity of the time delay factor $\rme^{-\rmi\Omega\tau}$.

In addition to giving zero--pole--gain expansions of the amplifier gain, reflectivity, and transmissivity, we shall expand the decay and gain rates to leading order via Taylor approximation, which assumes these rates are small relative to the cavity free spectral range.
For the decay rates, the Taylor approximation is
\begin{equation}
    r_j \simeq 1 - \kappa_j \fund{\tau}
    \label{eq:r approx}
\end{equation}
for $r_j \in \{r_1, r_2, \xtal{t}, r\}$.
If $r_j^2 + t_j^2 = 1$, which we will generally assume throughout, then $t_{1,2} \simeq \sqrt{2\kappa_{1,2} \fund{\tau}}$.
For small coupling rates, we can again write a Taylor approximation
\begin{subequations}
\label{eq:Gamma approx}
\begin{align}
    \cosh{\Gamma} &\simeq 1 \\
    \sinh{\Gamma} &\simeq \Gamma = g\fund{\tau}.
\end{align}
\end{subequations}
In some cases, expanding to leading order will require taking the Taylor approximations in \cref{eq:r approx,eq:Gamma approx} to quadratic order.

The choice of leading-order Taylor approximation is, in some sense, post hoc: we could have chosen to expand the rates by $[1/1]$ Padé approximation or by a Taylor approximation of higher order, but it is the leading-order Taylor approximation that closely reproduces the prediction under the Collett--Gardiner formalism, including the central result that the maximal amount of squeezed noise variance that can be produced within the bandwidth of a resonant OPA is given approximately by the escape loss $1 - \eta$ (\cref{eq:V11 max}).
A more refined expansion procedure for the rates could give greater accuracy, but is likely of limited utility because it is not too onerous to retain the exact rate expressions in the OPA transfer functions, as we will see later.

\begin{table}
    \centering
    \sisetup{
        exponent-mode=input,
        exponent-thresholds=-3:4,
        retain-unity-mantissa=true,
    }
    \caption{
        Parameters for a resonant optical parametric amplifier similar to the device used in Advanced LIGO~\cite{2015NatSR...518052W,T1700104,T2000273}; the value for the crystal loss is an ad-hoc assumption.
    \label{tab:vopo}}
    \IfPackageLoadedTF{fontspec}{\addfontfeature{Numbers=Lining}}{}
    \begin{tblr}{
        colspec={
            llS%
        },
        row{odd} = {bg=azure8!45!white},
        row{even} = {bg=azure8!20!white},
        row{1} = {bg=azure3, fg=white, font=\sffamily, guard},
    }
        &
        Symbol &
        Value %
        \\
        Fundamental wavelength [\unit{nm}] &
            $\fund{\lambda}$ &
            1064 %
            \\
        Round trip travel time [\unit{\ns}] &
            $\fund{\tau}$ &
            1.2 %
            \\
        Mirror~1 transmission [---] &
            $T_1$ &
            12.5\% %
            \\
        Mirror~2 transmission [---] &
            $T_2$ &
            0.15\% %
            \\
        Crystal length [\unit{\mm}] &
            $\xtal{l}$ &
            10 %
            \\
        Crystal loss [\unit{\cm^{-1}}] &
            $\mathcal{A}$ &
            0.01\% %
            \\
        Normalized nonlinear coupling [---] &
            $\Gamma/\Gamma_{\star}$ &
            0.80 %
            \\
    \end{tblr}
\end{table}

\section{Frequency-domain relations for sidebands and quadratures}
\label{sec:io}

We now compute the salient input--output relations (i.e., the transfer functions or frequency responses) that relate the optical fields injected into, circulating in, and exiting the amplifier.
We do so first in terms of sidebands, and then also in terms of quadratures.
Our two main goals will be to compute the amplifier's internal gain, which determines much of its behavior including its transmission properties, and the action of the amplifier on fields reflected from its overcoupled mirror (mirror~1 in \cref{fig:vopo}), which is where squeezed vacuum is produced.

Throughout the computation, we set the static detuning $\fund{\Phi}$ and the phase mismatch $\Delta k$ to zero.
In doing so, we are describing the amplifier's operation when it is perfectly on resonance, and we are not treating the mechanisms that give rise to, e.g., internal rotations of the squeezed and antisqueezed quadratures.
(We will, however, sometimes numerically evaluate the frequency responses at slightly nonzero pump phase to illustrate the cross-coupling of quadrature measurements in such a scenario.)
Such rotations can limit the signal-to-noise ratio enhancement that the amplifier can deliver in metrological applications like coherent interferometry.
The relations that result from our work can, like previous input--output relations, be used to estimate the cross-spectral density matrix of the squeezed vacuum state that results when the amplifier is seeded at each port by unsqueezed vacuum (in fact we do so in \cref{sec:noise}); it is then possible to estimate the effective squeezed variance due to external phase noise by including the noise in an rms sense~\cite{Aoki:2005gqw}.

Following previous work~\cite{McCuller:2021mbn}, we denote vectors of upper sidebands and their lower conjugate sidebands like $\symbf{a} = \begin{bmatrix*}[l] a_+ \\ a_-^\dagger \end{bmatrix*}$, and matrix relations between them with symbols like $\symsf{B}$.
We denote the corresponding vector of the cosine and sine quadratures like $\mathbb{a} = \begin{bmatrix} a_\text{c} \\ a_\text{s} \end{bmatrix}$, and the corresponding matrix relation like $\mathbb{B}$.
Quadratures are related to sidebands by
\begin{equation}
    \mathbb{a} = \symsf{A}\?\symbf{a}
\end{equation}
where
\begin{equation}
    \symsf{A} = \frac{1}{\sqrt{2}} \begin{bmatrix*}[r] 1 & 1 \\ -\rmi & +\rmi \end{bmatrix*}.
\end{equation}
Thus, a matrix relation $\symsf{B}$ of sidebands can be converted into a relation $\mathbb{B}$ of quadratures via
\begin{equation}
    \mathbb{B} = \symsf{A}\?\symsf{B}\?\symsf{A}^\dagger.
\end{equation}
We will examine the amplifier gain functions first (\cref{sec:G}), which are strictly proportional to the transmission functions (\cref{sec:T}).
We will then examine the reflection functions (\cref{sec:R}).

\begin{figure}[t]
  \centering
  \includegraphics[width=0.5\textwidth]{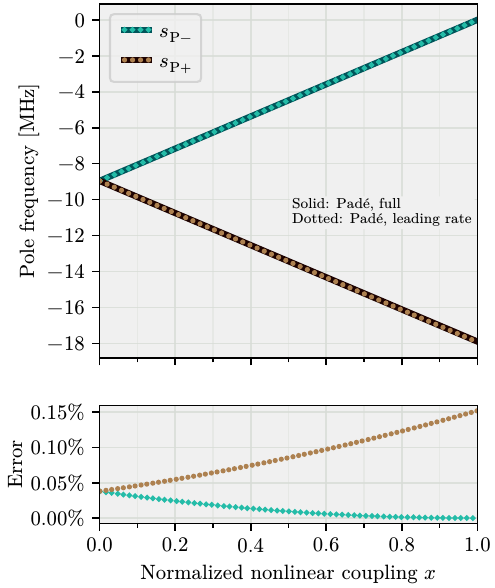}
  \caption{The cavity poles $s_{\text{P}\pm}$ in their full (\cref{eq:sP}) and approximate (\cref{eq:sP ll}) expressions, using the parameters in \cref{tab:vopo} but allowing the nonlinear coupling to vary.
  Because $s_{\text{P}-} \rightarrow 0$ as $x \rightarrow 1$ in both the full and approximate expressions, the error in the approximation also approaches 0.}
  \label{fig:poles}
\end{figure}

\begin{widetext}
\subsection{Intracavity gain functions}
\label{sec:G}

We met several gain functions in \cref{sec:fields}.
In addition to the cavity's single-pass gain (or open-loop gain) $\symsf{G}(s)$, we also defined the internal gain $\underline{\symsf{G}}(s)$ and the closed-loop gain $\overline{\symsf{G}}(s)$.\footnote{%
In a feedback control context, $\underline{\symsf{G}}$ may also be called the sensitivity function or the loop-suppressed gain.
}
Both of $\underline{\symsf{G}}$ and $\overline{\symsf{G}}$ characterize the cavity enhancement of the amplifier.
We compute frequency-domain expressions, in both the sideband and quadrature bases, for the cavity's internal gain \cref{sec:Gsupp} and the closed-loop gain in \cref{sec:Gc}.

\subsubsection{Internal gain}
\label{sec:Gsupp}

By referring to \cref{eq:G} and again defining $r = r_1 r_2 \xtal{t}$, we can write the internal gain as
\begin{equation}
    \underline{\symsf{G}}(s) = \left(\symsf{1} - \symsf{G}(s)\right)^{-1} = \frac{1}{1 - 2r\rme^{-s\fund{\tau}}\cosh{\Gamma} + r^2 \rme^{-2s\fund{\tau}}}
    \begin{bmatrix}
        1 - r\rme^{-s\fund{\tau}}\cosh{\Gamma} & r \rme^{-s\fund{\tau}}\rme^{-2\rmi\pump{\psi}} \sinh{\Gamma} \\
        r \rme^{-s\fund{\tau}}\rme^{+2\rmi\pump{\psi}} \sinh{\Gamma} & 1 - r\rme^{-s\fund{\tau}}\cosh{\Gamma} \\
    \end{bmatrix}.
    \label{eq:Gsupp}
\end{equation}

To begin examining this matrix, we define an amplified finesse
\begin{equation}
    \mathcal{F}_\Gamma = \frac{\pi}{\sqrt{1 - 2r\cosh{\Gamma} + r^2}},
    \label{eq:finesse}
\end{equation}
which coincides with the bare cavity finesse $\mathcal{F}_0 = \pi/(1-r)$ when $\Gamma = 0$.
(Other definitions of cavity finesse exist in the literature; our definition is chosen because it sets the scale of the internal gain, as we will see shortly.)
Applying the leading-order expansions described in \cref{sec:approximations}, we find
\begin{equation}
    \mathcal{F}_\Gamma \simeq \frac{\pi/\fund{\tau}}{\sqrt{\kappa^2 - g^2}}
    \label{eq:finesse ll}
\end{equation}
and hence $\mathcal{F}_0 \simeq \pi/\kappa\fund{\tau}$.
Unless otherwise specified, the symbol $\simeq$ refers to this kind of leading-order approximation.
We will make comparisons between the exact and approximate expressions for $\mathcal{F}_\Gamma$ in \cref{sec:gains}.

We now undertake the Padé expansion of $\underline{\symsf{G}}$ by the strategy in \cref{eq:Padé}. 
We note that the entries of $\underline{\symsf{G}}$ contain factors up to $\left(r\rme^{-s\fund{\tau}}\right)^2$ in both the numerator and denominator; therefore, in making this Padé expansion, we expect to arrive at a series of rational expressions of second order.
First, for the prefactor of $\underline{\symsf{G}}$, we find
\begin{equation}
\frac{1}{1 - 2r\rme^{-s\fund{\tau}}\cosh{\Gamma} + r^2 \rme^{-2s\fund{\tau}}}
        \xrightarrow{\text{Padé}} \frac{\mathcal{F}_\Gamma^2}{\pi^2} \times \frac{\left(1 + s\fund{\tau}/2\right)^2}{\left(1 - s/s_{\text{P}+}\right)\left(1 - s/s_{\text{P}-}\right)}
\end{equation}
with the poles
\begin{align}
    s_{\text{P}\pm} &= \frac{2}{\fund{\tau}} \times \frac{r^2 \mp 2r\sinh{\Gamma} - 1}{r^2 + 2r \cosh{\Gamma} + 1}
    \label{eq:sP} \\
    &\simeq -\kappa \mp g.
    \label{eq:sP ll}
\end{align}
\end{widetext}
Since the amplifier is operated below threshold and all the rates are positive by definition, the poles $s_{\text{P}\pm}$ are always in the left half of the complex plane ($\real{s} < 0$).
For the amplifier in \cref{tab:vopo}, one finds $s_{\text{P}+}/2\pi \approx \qty{-16.1}{\MHz}$, with the approximation yielding a fractional error of \qty{0.12}{\%}.
Meanwhile, $s_{\text{P}-}/2\pi \approx \qty{-1.79}{\MHz}$, with an error of \qty{15}{ppm}.
A plot of the exact and approximate expressions for these poles is shown in \cref{fig:poles} as a function of nonlinear interaction strength.

We continue on to make a Padé expansion of the rest of $\underline{\symsf{G}}$, and we find after some factorization that we can write it as
\begin{equation}
    \underline{\symsf{G}}(s) \xrightarrow{\text{Padé}} \frac{\left(1 + s\fund{\tau}/2\right)\times\underline{\symsf{G}}(0)}{\left(1 - s/s_{\text{P}+}\right)\left(1 - s/s_{\text{P}-}\right)} \odot
    \begin{bmatrix}
        1 - s/s_{\text{T}} & 1 - s\fund{\tau}/2 \\
        1 - s\fund{\tau}/2 & 1 - s/s_{\text{T}}
    \end{bmatrix}
    \label{eq:Gsupp zpk}
\end{equation}
where $\odot$ is the elementwise (Hadamard) matrix product.
The dc portion of $\underline{\symsf{G}}$ is
\begin{align}
    \underline{\symsf{G}}(0) &= \frac{\mathcal{F}_\Gamma^2}{\pi^2} 
    \begin{bmatrix}
        1 - r\cosh{\Gamma} & r\rme^{-2\rmi\pump{\psi}}\sinh{\Gamma} \\
        r\rme^{+2\rmi\pump{\psi}}\sinh{\Gamma} & 1 - r\cosh{\Gamma}
    \end{bmatrix} \label{eq:Gsupp dc} \\
    &\simeq
    \frac{1/\fund{\tau}}{\kappa^2 - g^2} \begin{bmatrix}
        \kappa & g \rme^{-2\rmi\pump{\psi}} \\
        g \rme^{+2\rmi\pump{\psi}} & \kappa \\
    \end{bmatrix}
    \label{eq:Gsupp zpk ll dc}
\end{align}
and we have defined the zero
\begin{align}
    s_{\text{T}} &= \frac{2}{\fund{\tau}} \times \frac{r\cosh{\Gamma} - 1}{r\cosh{\Gamma}+1} \\
    &\simeq -\kappa. \label{eq:sT ll}
\end{align}
From \cref{tab:vopo}, the example amplifier has $s_{\text{T}}/2\pi \approx \qty{-8.8}{\MHz}$ from the exact expression.
The approximation is \qty{2.2}{\%} larger in magnitude, and one can see from the form in \cref{eq:sT ll} that unlike the exact expression, the approximation does not shift under the presence of the nonlinear interaction.
In addition to this zero, all elements of $\underline{\symsf{G}}$ contain a left half-plane zero at $-2/\fund{\tau}$, and the off-diagonal elements also contain a right half-plane zero at $+2/\fund{\tau}$.
The impact of these zeros at Fourier frequencies far below the free spectral range $2\pi/\fund{\tau}$ is noticeable but modest: for sidebands at $s = \rmi\kappa$, for instance, each zero imparts a phase shift of about $\pm\qty{8}{\degree}$.

This Padé expansion and leading-order approximation already gives us some insight into the evolution of a sideband at frequency $\fund{\omega}\pm\Omega$ injected into the amplifier.
From the calculations above, we see that within the amplifier bandwidth ($|\Omega| \lesssim \left|s_{\text{P}-}\right| \simeq |\kappa - g|$), the effect of $\underline{\symsf{G}}$ is to generate the complementary sideband at $\fund{\omega}\mp\Omega$ with a complex amplitude $\simeq (g/\kappa)\rme^{\mp2\rmi\pump{\psi}}$ relative to the injected sideband; this is shown in \cref{fig:RT} using amplifier parameters from \cref{tab:vopo}.
Near and above the amplifier bandwidth, the sidebands are not fully resonant in the cavity and so the cavity enhancement of the nonlinear interaction is reduced.
Thus, the strength of the complementary sideband generation is diminished relative to the circulating injected sideband.
When the nonlinear interaction is turned off ($\Gamma \rightarrow 0$, or $g \rightarrow 0$), no complementary sideband is generated, and the poles $s_{\text{P}\pm}$ and the zero $s_{\text{T}}$ coalesce to $(2/\fund{\tau})(r - 1)/(r + 1) \simeq -\kappa$, so that the gain of the injected sideband has a single-pole form, consistent with the behavior of a passive cavity.
Conversely, at threshold, $s_{\text{P}+} \rightarrow (2/\fund{\tau})(r^2 - 1)/(r^2 + 1) \simeq -2\kappa$, and $s_{\text{P}-} \rightarrow 0$.

This basic picture of the amplifier's gain can also be deduced from the Hamiltonian formalism. %
Indeed, in \cref{sec:cg} we compare how an approximation to $\underline{\symsf{G}}$ can be derived via mean-field equations of motion, which are the same equations of motion that result from the Hamiltonian formalism. %
We find that the mean-field procedure gives expressions identical to our approximate expressions \cref{eq:Gsupp zpk,eq:Gsupp zpk ll dc,eq:sT ll}, except our expressions contain zeros at $\pm2/\fund{\tau}$ that are absent in the mean-field approximation.
This is because the mean-field approximation does not consider the free spectral range of the cavity, except as an overall scale factor that sets the coupling and decay rates.

\begin{widetext}
Having examined the internal gain in the sideband basis, we now examine its action on optical quadratures.
We can transform into the quadrature basis by writing
\begin{multline}
    \underline{\mathbb{G}}(s) = \symsf{A}\?\underline{\symsf{G}}(s)\?\symsf{A}^\dagger = 
\frac{1}{1 - 2r\rme^{-s\fund{\tau}}\cosh{\Gamma} + r^2 \rme^{-2s\fund{\tau}}} \\
\times \begin{bmatrix}
1 + r\rme^{-s\fund{\tau}}\left(\cos {2 \pump{\psi}} \sinh {\Gamma} - \cosh {\Gamma}\right) &
- r\rme^{- s \tau_{a}} \sin {2 \pump{\psi}} \sinh {\Gamma}\\
- r\rme^{- s \tau_{a}} \sin {2 \pump{\psi}} \sinh {\Gamma} &
1 - r\rme^{-s\fund{\tau}}\left(\cos {2 \pump{\psi}} \sinh {\Gamma} + \cosh {\Gamma}\right) \\
\end{bmatrix},
    \label{eq:Gsupp q}
\end{multline}
and then Padé expand as before:
\begin{equation}
    \underline{\mathbb{G}}(s) \xrightarrow{\text{Padé}} \frac{(1 + s\fund{\tau}/2)\times\underline{\mathbb{G}}(0)}{\left(1 - s/s_{\text{P}+}\right)\left(1 - s/s_{\text{P}-}\right)}
    \odot \begin{bmatrix} 1 - s/s_{\text{Tq}-} & 1 - s\fund{\tau}/2 \\ 1 - s\fund{\tau}/2 & 1 - s/s_{\text{Tq}+} \end{bmatrix},
    \label{eq:Gsupp q zpk}
\end{equation}
with
\begin{align}
    \underline{\mathbb{G}}(0) &= \frac{\mathcal{F}_\Gamma^2}{\pi^2} \begin{bmatrix}
1 + r \cos {2 \pump{\psi}} \sinh {\Gamma} - r \cosh {\Gamma} & - r \sin {2 \pump{\psi}} \sinh {\Gamma}\\- r \sin {2 \pump{\psi}} \sinh {\Gamma} & 1 - r \cos {2 \pump{\psi}} \sinh {\Gamma} - r \cosh {\Gamma} \end{bmatrix} \\
    &\simeq \frac{1/\fund{\tau}}{\kappa^2 - g^2} \begin{bmatrix} \kappa + g\cos{2\pump{\psi}} & -g\sin{2\pump{\psi}} \\ -g\sin{2\pump{\psi}} & \kappa - g\cos{2\pump{\psi}}
    \end{bmatrix}
\end{align}
and
\begin{align}
    s_{\text{Tq}\mp} &= \frac{2}{\fund{\tau}} \times \frac{\mp r\cos{2\pump{\psi}}\sinh{\Gamma} + r\cosh{\Gamma} -1 }{\mp r\cos{2\pump{\psi}}\sinh{\Gamma} + r\cosh{\Gamma} + 1} \\
    &\simeq -\kappa \mp g \cos{2\pump{\psi}}.%
\end{align}
\end{widetext}

Evidently, when $\pump{\psi} = 0$, fluctuations in the cosine quadrature experience an amplitude gain of $\left. 1 \middle/\left(1 - r \rme^{+\Gamma}\right)\right. \simeq 1/\fund{\tau}(\kappa - g)$, while fluctuations in the sine quadrature experience a gain of $\left. 1 \middle/\left(1 - r \rme^{-\Gamma}\right)\right. \simeq 1/\fund{\tau}(\kappa+g)$.
In a passive cavity, both quadratures would experience an identical gain of $1/(1-r) \simeq 1/\fund{\tau}\kappa$, showing the phase-sensitive action of the amplifier.
Indeed, at threshold $x \rightarrow 1$ one finds the intracavity fluctuations in the sine quadrature are deamplified relative to a passive cavity by an amplitude factor $1/(1 + r) \simeq 2$, which is consistent with early quantum optics calculations: one does not a high degree of squeezing inside the amplifier cavity; rather, it is the interference of fields leaving the external side of mirror 1 where squeezing is found~\cite{1981OptCo..39..401M,1984PhRvA..29..408Y}.
The bandwidths of the gains in each quadrature differ: for $\pump{\psi} = 0$, the zeros $s_{\text{Tq}\mp}$ coalesce to $s_{\text{P}\pm}$, and hence the overall frequency response for the gain in the cosine quadrature has only a single pole at $s_{\text{P}-} \simeq -\kappa + g$.
For the same reason, the sine quadrature has only a single pole at $s_{\text{P}+} \simeq -\kappa - g$, showing that the deamplification occurs over a wider bandwidth than the amplification.
However, one can see from the leading-order rate approximation that both quadratures have an identical gain--bandwidth product of $1/\fund{\tau}$.

\begin{figure}[t]
  \centering
  \includegraphics[width=0.5\textwidth]{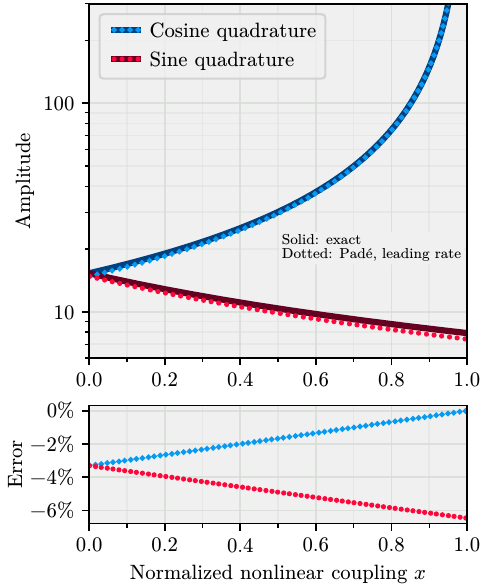}
  \caption{The exact and approximate forms of the dc internal gain $\underline{\mathbb{G}}(0)$.
  The upper element $\underline{\mathbb{G}}_{11}(0)$ approaches $\infty$ as $x \rightarrow 1$, both in the exact and approximate forms, and the approximation error goes to 0.}
  \label{fig:dc gains}
\end{figure}

The sideband and quadrature frequency responses of the internal gain for a resonant optical parametric amplifier using parameters from \cref{tab:vopo} are plotted in \cref{fig:RT}, showing both the exact frequency responses of \cref{eq:Gsupp,eq:Gsupp q} and the zero--pole--gain Padé expansions \cref{eq:Gsupp zpk,eq:Gsupp q zpk} with their leading-order rate approximations.
Here we have set $\pump{\psi} = \qty{3}{\milli\radian}$ to illustrate the coupling of the two quadratures that results when examining the amplifier response when slightly mistuned from the pump phase.
The greatest inaccuracy is in the approximation of $\underline{\mathbb{G}}_{22}$, which results in a dc fractional error of \qty{6}{\%} for the amplifier considered here.
\cref{fig:dc gains} shows $\underline{\mathbb{G}}_{11}(0)$ and $\underline{\mathbb{G}}_{22}(0)$ as a function of interaction strength, in their exact and approximate forms.\footnote{%
Note that if a more accurate approximation to $\underline{\mathbb{G}}(0)$ is desired, one can make a more precise expansion of the quantity $r\rme^{\pm\Gamma} = \rme^{(\kappa\pm{}g)\tau}$ rather than perform the first-order Taylor approximations in \cref{eq:r approx,eq:Gamma approx}.
For the $[1/1]$ Padé approximation, $\left.1\middle/\left(1 - r\rme^{\pm\Gamma}\right)\right. \xrightarrow{\text{Padé}} \left(1 + (\kappa\pm{}g)\fund{\tau}/2\right)/\fund{\tau}(\kappa\pm{}g)$; this reduces the fractional approximation error in $\underline{\mathbb{G}}_{22}(0)$ to \qty{0.13}{\%}.%
}

\subsubsection{Closed loop gain}
\label{sec:Gc}

The closed-loop gain of the cavity is 
\begin{align}
    \overline{\symsf{G}}(s) &= \symsf{G}(s) (\symsf{1} - \symsf{G}(s))^{-1} \\
    &=  \frac{r \rme^{-s\fund{\tau}}}{1 - 2r\rme^{-s\fund{\tau}}\cosh{\Gamma} + r^2 \rme^{-2s\fund{\tau}}} \nonumber \\
    &\hphantom{=}\qquad\times
    \begin{bmatrix} \cosh {\Gamma} - r\rme^{-s\fund{\tau}} &\rme^{-2\rmi\pump{\psi}} \sinh {\Gamma}\\
    \rme^{+2\rmi\pump{\psi}} \sinh {\Gamma} & \cosh{\Gamma} - r\rme^{-s\fund{\tau}} \end{bmatrix}.
    \label{eq:Gc}
\end{align}
Because of the amplifier topology considered here, we will not need to examine $\overline{\symsf{G}}$ in detail, although we will need it as an ingredient in the reflection transfer functions in \cref{sec:R}.
Explicit computation of Padé expansions and leading-order rate approximations for $\overline{\symsf{G}}$ and $\overline{\mathbb{G}}$ is done in \cref{sec:cloop}.
To leading order in the rates, $\overline{\symsf{G}} \simeq \underline{\symsf{G}}$, and $\overline{\mathbb{G}} \simeq \underline{\mathbb{G}}$, which was anticipated at dc and in the case $r = r_1$~\cite{Ganapathy:2022hgu}.

\subsection{Transmission functions}
\label{sec:T}

Referring to \cref{eq:a1prime}, we define the transmission of the sidebands $\symbf{a}^{(2)} \mapsto \symbf{a}^{(1')}$ as
\begin{equation}
    \symsf{T}(s) = t_1 t_2 \underline{\symsf{G}}(s).
    \label{eq:T}
\end{equation}
This expression for $\symsf{T}$ is appropriate for amplifiers with topology as shown in \cref{fig:vopo};
a different arrangement of ports can instead result in the transmission functions being proportional to $\overline{\symsf{G}}$~\cite[Appx.~C]{Ganapathy:2022hgu}.

The transmission of sidebands entering via the crystal loss and exiting mirror~1\,---\,i.e., $\symbf{a}^{(\text{X})} \mapsto \symbf{a}^{(1')}$\,---\,is
\begin{equation}
    \symsf{T}^{(\text{X})}(s) = t_1 r_2 \xtal{r} \underline{\symsf{G}}(s).
    \label{eq:TX}
\end{equation}
In defining both $\symsf{T}$ and $\symsf{T}^{(\text{X})}$ we have omitted the various propagation matrices $\symsf{L}_j$ in \cref{eq:a1prime}, as they can be accounted for by transformations of the ingoing and outgoing sideband vectors.
Since $\symsf{T}$ and $\symsf{T}^{(\text{X})}$ are simply proportional to the internal gain, we will not further examine their functional forms except to note the leading-order rate approximations $t_1 t_2 \simeq 2\sqrt{\kappa_1 \kappa_2}\fund{\tau}$ and $t_1 r_2 \xtal{r} \simeq 2\sqrt{\kappa_1 \xtal{\kappa}}\fund{\tau}$.

\subsection{Reflection functions}
\label{sec:R}

We consider two reflection functions: the reflection of fields at the external side of mirror~1, which generates the squeezed vacuum (\cref{sec:R1}), and the reflection of fields at the external side of mirror~2 (\cref{sec:R2}), which is used for controls and diagnostics.
These two functions have the same mathematical form, but they have different phenomenology due to the different decay rates, with $\kappa_2 \ll \kappa_1$.

\begin{figure*}[t]
    \centering
    \includegraphics[width=0.5\textwidth]{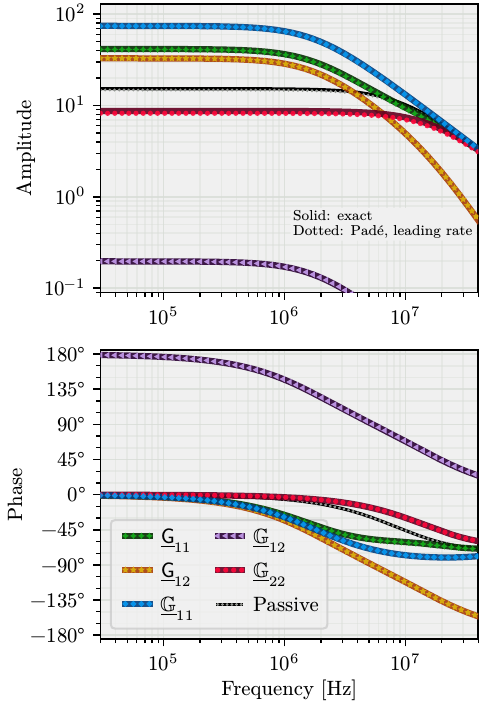}%
    \includegraphics[width=0.5\textwidth]{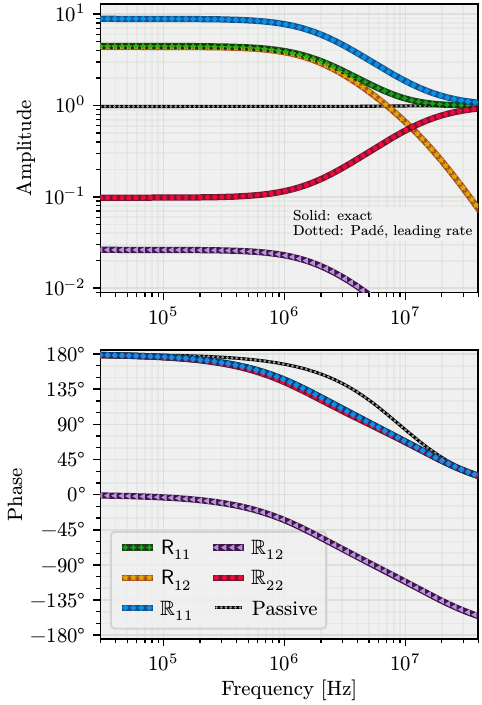}
    \caption{Input--output relations for the internal gain (left; \cref{sec:Gsupp}) and reflection (right; \cref{sec:R}) for a resonant optical parametric amplifier with parameters given in \cref{tab:vopo}, showing the unique entries for both the sideband representations $\underline{\symsf{G}}$ and $\symsf{R}$ and the quadrature representations $\underline{\mathbb{G}}$ and $\mathbb{R}$.
    The relations are plotted at a small but nonzero pump phase ($\pump{\psi} = \qty{3}{\milli\radian}$) to exhibit cross-coupling of the quadratures.
    Solid lines show the exact expressions; dotted lines show the Padé expansions into zero--pole--gain form and with approximations for small decay and gain rates.
    The zeros at $s = -2/\fund{\tau}$ contribute nonnegligibly to the phase response of $\underline{\symsf{G}}$ and $\underline{\mathbb{G}}$ above \qty{10}{\MHz}.
    Also shown is the passive limit ($\Gamma \rightarrow 0$ and $g \rightarrow 0$), to which the diagonal elements of the sideband and quadrature relations coalesce; the off-diagonal elements go to $0$.}
    \label{fig:RT}
\end{figure*}

\subsubsection{Reflection from mirror~1 (squeezing port)}
\label{sec:R1}

From \cref{eq:a1prime}, the reflection $\symsf{R}$, which maps the sidebands $\symbf{a}^{(1)} \mapsto \symbf{a}^{(1')}$ at mirror~1 (\cref{fig:vopo}), is 
\begin{equation}
    \symsf{R}(s) = r_1 \symsf{1} - \frac{t_1^2}{r_1} \overline{\symsf{G}}(s).
    \label{eq:Rs}
\end{equation}
In the usual fashion with optical cavities, the reflection of the ingoing field is a superposition of a prompt reflection with amplitude $r_1$, and a component that enters and circulates inside the amplifier, hence being enhanced by the gain $\overline{\symsf{G}}$, before leaking out.
As with the transmission functions, we have not included any explicit appearances of the propagation matrices $\left\{\symsf{L}_j\right\}$, as these can be absorbed into the sidebands.

Having previously computed a full expression for $\overline{\symsf{G}}$ in \cref{eq:Gc}, it is trivial to write down a full expression for $\symsf{R}$ using \cref{eq:Rs}, but the resulting expression is not illuminating, so we omit it.
Instead, we proceed with the Padé expansion.
One can deduce from \cref{eq:Rs} that the entries of $\symsf{R}$, like the entries of $\overline{\symsf{G}}$, contain terms up to $(r\rme^{s\fund{\tau}})^2$ in their numerators and denominators.
We therefore anticipate the Padé expansion of the exponential will yield second-order rational functions.
\begin{widetext}%
Indeed, performing the expansion using \cref{eq:Padé} and then factoring into poles and zeros, we find
\begin{equation}
    \symsf{R}(s)
        \xrightarrow{\text{Padé}} \frac{\symsf{R}(0)}{\left(1 - s/s_{\text{P}+}\right)\left(1 - s/s_{\text{P}-}\right)} \odot
        \begin{bmatrix}
            (1 - s/s_{\text{R}+})(1 - s/s_{\text{R}-}) &
            (1 - s\fund{\tau}/2)(1 + s\fund{\tau}/2) \\
            (1 - s\fund{\tau}/2)(1 + s\fund{\tau}/2) &
            (1 - s/s_{\text{R}+})(1 - s/s_{\text{R}-}) \\
        \end{bmatrix},
\end{equation}
with $\odot$ again denoting the Hadamard product.
The dc values of $\symsf{R}$ are 
\begin{align}
    \symsf{R}(0) &= \frac{\mathcal{F}_\Gamma^2/\pi^2}{r_1}
    \begin{bmatrix}
        r t_1^2 \left(r - \cosh {\Gamma}\right) + r_1^2  (\pi/\mathcal{F}_\Gamma)^2 &
        - r t_1^2 \rme^{- 2\rmi\pump{\psi}} \sinh {\Gamma}\\
        - r t_1^2 \rme^{2\rmi\pump{\psi}} \sinh {\Gamma} &
        r t_1^2 \left(r - \cosh {\Gamma}\right) + r_1^2 (\pi/\mathcal{F}_\Gamma)^2
    \end{bmatrix} \\
    &\simeq \frac{1}{\kappa^2 - g^2} \begin{bmatrix}
        - g^2 - 2\kappa_1\kappa + \kappa^2 & - 2 g \kappa_1 \rme^{-2\rmi\pump{\psi}}\\
        - 2 g \kappa_1 \rme^{+2\rmi\pump{\psi}} & - g^2 - 2 \kappa_1\kappa + \kappa^2
    \end{bmatrix}.
\end{align}
The zeros $s_{\text{R}\pm}$ are\footnote{%
Here and in the expressions for the quadrature zeros $s_{\text{Rq}\pm\pm'}$ later, the term $r_1^2 + t_1^2$ appears in several places, although we have assumed in numerical simulation that it is equal to $1$; i.e., the reflective interface of mirror~1 is lossless.
Including loss at this interface would require the consideration of an additional vacuum sideband field similar to the crystal loss field $\symbf{a}^{(\text{X})}$.%
}
\begin{align}
    s_{\text{R}\pm} &= \frac{2}{\fund{\tau}} \times \frac{(r^2 - 1) r_1^2  + r^2 t_1^2 \pm r \sqrt{4 r_1^2  \left(r_1^2  + t_1^2\right) \sinh^2{\Gamma} + t_1^4 \cosh^2{\Gamma}}}{(r^2 + 1) r_1^2  + r^2 t_1^2 + r \left(2 r_1^2  + t_1^2 \right)\cosh{\Gamma}} \\
    &\simeq -\kappa + \kappa_1 \pm \sqrt{\kappa_1^2 + g^2}.
\end{align}
The upper components of $\symsf{R}$, both exactly and in the leading-order approximation to their Padé expansions, are plotted in \cref{fig:RT}.
Similar to the situation with $\underline{\symsf{G}}$, when a sideband at $\fund{\omega} \pm \Omega$ is reflected from mirror~1, a complementary sideband at $\fund{\omega} \mp \Omega$ is generated within the cavity bandwidth ($|\Omega| \lesssim |s_{\text{P}-}|$).
Both the reflected sideband and the conjugate of its complement have nearly the same complex amplitude, and both amplitudes are larger than the amplitude of the initial sideband.
Because of this, when transformed into the quadrature basis, one quadrature is amplified and the other is deamplified, which demonstrates the principle of squeezed vacuum generation by optical parametric amplification.

As with the gains, we can write down the reflection matrix in the quadrature basis:
We now compute the reflection matrix in the quadrature basis:
\begin{equation}
    \mathbb{R}(s) = \symsf{A}\?\symsf{R}(s)\?\symsf{A}^\dagger
\end{equation}
This, too, is not illuminating to write out, and we instead present its Padé expansion:
\begin{equation}
    \mathbb{R}(s) \xrightarrow{\text{Padé}} \frac{\mathbb{R}(0)}{\left(1 - s/s_{\text{P}+}\right)\left(1 - s/s_{\text{P}-}\right)} \odot
    \begin{bmatrix}
        (1 - s/s_{\text{Rq}++'})(1 - s/s_{\text{Rq}-+'}) &
        (1 - s\fund{\tau}/2)(1 + s\fund{\tau}/2) \\
        (1 - s\fund{\tau}/2)(1 + s\fund{\tau}/2) &
        (1 - s/s_{\text{Rq}+-'})(1 - s/s_{\text{Rq}--'})
    \end{bmatrix}
    \label{eq:R Padé}
\end{equation}
with dc values
\begin{align}
    \mathbb{R}(0) &=
    \frac{\mathcal{F}_\Gamma^2/\pi^2}{r_1 } \begin{bmatrix}
      r_1^2(\pi/\mathcal{F}_\Gamma)^2 + r t_1^2 (r -\cos {2 \pump{\psi}} \sinh {\Gamma} - \cosh {\Gamma}) &
      r t_1^2 \sin {2 \pump{\psi}} \sinh {\Gamma} \\
      r t_1^2 \sin {2 \pump{\psi}} \sinh {\Gamma} &
        r_1^2(\pi/\mathcal{F}_\Gamma)^2 + r t_1^2 (r + \cos {2 \pump{\psi}} \sinh {\Gamma} - \cosh {\Gamma})
    \end{bmatrix} \\
    &\simeq \frac{1}{\kappa^2 - g^2}
    \begin{bmatrix}
        \left(\kappa - g\right)\left(\kappa + g\right) - 2\kappa_1 \left(\kappa + g\cos{2\pump{\psi}}\right) & 2g\kappa_1 \sin {2 \pump{\psi}}\\ 2g\kappa_1 \sin {2 \pump{\psi}} & \left(\kappa - g\right)\left(\kappa + g\right) - 2\kappa_1 \left(\kappa - g\cos{2\pump{\psi}}\right)
    \end{bmatrix} \\
    &\xrightarrow{\pump{\psi} \rightarrow 0} \begin{bmatrix}
        \frac{\kappa - 2\kappa_1 - g}{\kappa - g} & 0 \\
        0 & \frac{\kappa - 2\kappa_1 + g}{\kappa + g}
    \end{bmatrix}.
    \label{eq:Rq zpk dc ll}
\end{align}
As we anticipated in the discussion above, when $\pump{\psi} = 0$, and in the usual circumstance when $\kappa_1 / \kappa$ is nearly 1, the cosine quadrature is amplified and the sine quadrature is deamplified, so under our conventions, we will find squeezed vacuum in the sine quadrature of the outgoing field $\mathbb{a}^{(1')}$ when the ingoing field $\mathbb{a}^{(1)}$ is initialized in an unsqueezed vacuum state.

Continuing on to examine the frequency dependence, the four zeros $s_{\text{R}\pm\pm'}$ in \cref{eq:R Padé} are solutions to the equation
\begin{equation}
    0 = a_{\pm'} (s_{\text{Rq}\pm\pm'} \fund{\tau})^2 + b s_{\text{Rq}\pm\pm'} \fund{\tau} + c_{\pm'}
    \label{eq:sRq quadratic}
\end{equation}
with
\begin{subequations}
    \label{eq:sRq abc}
\begin{align}
    a_{\pm'} &= r^2 \left(r_1^2 + t_1^2\right) + r \left(\pm' t_1^2 \cos {2 \psi_{b}} \sinh {\Gamma} + \left(2 r_1^2 + t_1^2\right) \cosh {\Gamma}\right) + r_1^2 \\
    b &= 4 \left(r_1^2 - r^2 \left(r_1^2 + t_1^2\right)\right) \\
    c_{\pm'} &= 4 r^2 \left(r_1^2 + t_1^2\right) - 4 r \left(\pm' t_1^2 \cos {2 \psi_{b}} \sinh {\Gamma} + \left(2 r_1^2 + t_1^2\right) \cosh {\Gamma}\right) + 4 r_1^2,
\end{align}
\end{subequations}
which yields, in the leading-order rate approximation,
\begin{align}
    s_{\text{Rq}\pm\pm'} &\simeq -\kappa + \kappa_1 \pm \sqrt{\kappa_1^2 \pm' 2g\kappa_1 \cos{2\pump{\psi}} + g^2} \\
    &\xrightarrow{\pump{\psi}\rightarrow{}0} -\kappa + \kappa_1 \pm (\kappa_1 \pm' g) \qquad\qquad \text{(if $\kappa_1 \ge g$)}.
\end{align}
\end{widetext}

It is worth remarking on the specific functional form of the zeros $s_{\text{Rq}\pm\pm'}$ when $\pump{\psi} = 0$.
For the cosine quadrature, $\pm' = +$ and so the zero $s_{\text{Rq}-+} = -\kappa - g$ cancels the pole $s_{\text{P}+} = -\kappa - g$.
The frequency response in this quadrature therefore has the single pole $s_{\text{P}-} = -\kappa + g$ and the single zero $s_{\text{Rq}++} = -\kappa + 2\kappa_1 + g$.
In the sine quadrature one reaches the complementary conclusion that the frequency response has the single pole $s_{\text{P}+} = -\kappa - g$ and the single zero $s_{\text{Rq}+-} = -\kappa + 2\kappa_1 - g$.
When $\kappa_1 / \kappa$ is close to unity, as required to generate a high degree of squeezing, the zeros for both quadratures are in the right half-plane, which can be seen from the Bode plot in \cref{fig:RT}.
This is the expectation for an overcoupled cavity.

\subsubsection{Reflection from mirror~2 (control port)}
\label{sec:R2}

As we remarked earlier, it is desirable to inject coherent sidebands for control purposes or for diagnostics.
Often, this injection consists of a single sideband detuned by megahertz frequencies above the fundamental carrier frequency, and the control relies on the amplifier's ability to generate the complementary sideband below the carrier~\cite{Vahlbruch:2006df,Chelkowski:2007teo,Dwyer:2013tth,2016Optic...3..682O}.
In the model amplifier considered in \cref{fig:vopo}, the relevant injection occurs at mirror~2 and information for feedback control is available by monitoring the reflected coherent field from mirror~2.
The frequency response of this reflection is therefore relevant to examine, primarily in its sideband representation.

From \cref{eq:a2prime}, it is evident that the reflection $\symsf{R}^{(2)}$, which maps $\symbf{a}^{(2)} \mapsto \symbf{a}^{(2')}$ at mirror~2, is formally equivalent to the reflection $\symsf{R}$, which maps $\symbf{a}^{(1)} \mapsto \symbf{a}^{(1')}$ at mirror~1 (which we might retroactively notate as $\symsf{R}^{(1)}$).
The mathematical analysis for $\symsf{R}^{(1)}$ in \cref{sec:R1} therefore carries over directly to $\symsf{R}^{(2)}$ by exchange of subscripts $1 \leftrightarrow 2$ in the reflectivities and decay rates.
We will not produce full formulae for $\symsf{R}^{(2)}$, since they can be deduced from the previous formulae for $\symsf{R}^{(1)}$.
However, we will note the following important phenomenological differences between $\symsf{R}^{(1)}$ and $\symsf{R}^{(2)}$ in the leading-order rate approximation.

By design, mirror~1 strongly overcouples the field $\symbf{a}^{(1)}$ to the cavity; i.e, $\kappa_1 / \kappa \approx 1$, and we have described in \cref{sec:R1} and \cref{fig:RT} the typical form of $\symsf{R}^{(1)}$ and $\mathbb{R}^{(1)}$ therefrom.
Conversely, mirror~2 strongly undercouples the field $\symbf{a}^{(2)}$, since $\kappa_2 / \kappa \ll 1$.
\begin{widetext}
In the leading-order rate approximation if one additionally takes $\kappa \pm \kappa_2 \rightarrow \kappa$, the reflection from mirror~2 becomes 
\begin{equation}
    \symsf{R}^{(2)}(s) \simeq \begin{bmatrix}
        1 & -\frac{2 g \kappa_2 \rme^{-2\rmi\pump{\psi}}}{\kappa^2 - g^2} \\
        -\frac{2 g \kappa_2 \rme^{+2\rmi\pump{\psi}}}{\kappa^2 - g^2} & 1 \\
    \end{bmatrix} \odot
    \begin{bmatrix}
        1 & \frac{(1 - s\fund{\tau}/2)(1 + s\fund{\tau}/2)}{(1 - s/s_{\text{P}+})(1 - s/s_{\text{P}-})} \\
        \frac{(1 - s\fund{\tau}/2)(1 + s\fund{\tau}/2)}{(1 - s/s_{\text{P}+})(1 - s/s_{\text{P}-})} & 1 \\
    \end{bmatrix}.
\end{equation}
\end{widetext}
This shows that a sideband injected at $\fund{\omega} \pm \Omega$ is almost perfectly reflected from mirror~2, with (under this approximation) no effect from the cavity gain.
So long as $g \ne 0$ and $\kappa_2 \ne 0$, a small complementary sideband at $\fund{\omega} \mp \Omega$ is generated, with amplitude proportional to $\left(\mathcal{F}_\Gamma\middle/\pi\right)^2 g\kappa_2$ at dc.
This complementary sideband experiences both poles of the cavity's internal gain, as well as two high-frequency zeros associated with the cavity's free spectral range.

\section{Amplifier figures of merit}
\label{sec:gains}

This section examines some quantities that are used to characterize the performance of a resonant parametric amplifier, especially its gain and its escape efficiency.
Many of the quantities are measurements of the amplifier's dc response, but information is also available from the ac response and we will also examine those aspects here.

In \cref{eq:finesse} above, we derived an amplified finesse that characterizes the amplifier's internal gain.
Recalling the definition of threshold gain $\Gamma_{\star} = -\ln{r}$, it follows from \cref{eq:finesse} that the amplified finesse becomes infinite at threshold:
\begin{equation}
    \mathcal{F}_\Gamma \xrightarrow{\Gamma \rightarrow \Gamma_{\star}} \infty.
\end{equation}
This similarly holds for the leading-order rate approximation \cref{eq:finesse ll} when $g \rightarrow \kappa$.
Additionally, as the amplifier approaches threshold, its bandwidth approaches 0;
this can be seen by inspecting, e.g., the lowest-frequency pole $s_{\text{P}-} \simeq -\kappa + g$ (\cref{eq:sP ll}), which appears in the dynamics of the amplifier's internal gain $\underline{\symsf{G}}$. 

\begin{figure*}[t]
    \centering
    \includegraphics[width=0.5\textwidth]{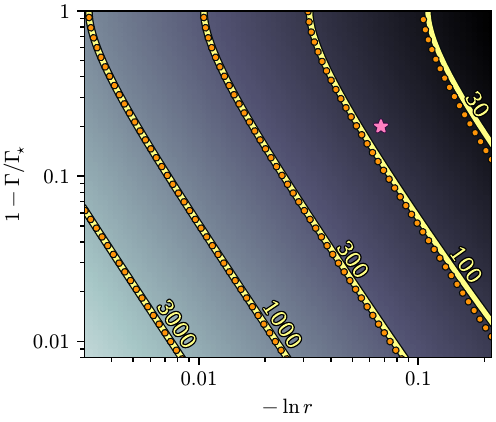}%
    \includegraphics[width=0.5\textwidth]{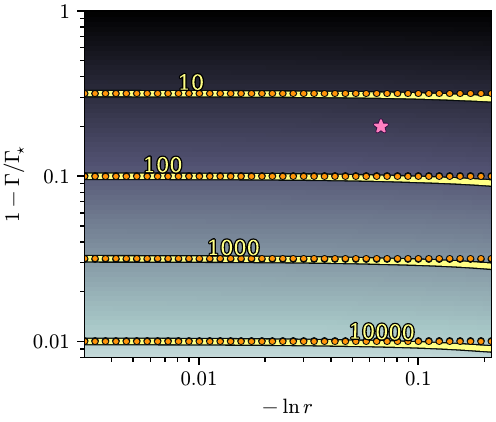}
    \caption{\emph{Left:} amplified finesse in its exact expression (solid, \cref{eq:finesse}) and in the leading-order rate approximation (dotted; \cref{eq:finesse ll}). \emph{Right:} nonlinear gain in its exact expression (solid; \cref{eq:Gnl}) and in the leading-order rate approximation (dotted; \cref{eq:Gnl ll}).
    The quantities are plotted as a function of both $-\ln{r} = \kappa\fund{\tau}$ (the decay rate relative to the cavity free spectral range) and the nonlinear interaction $\Gamma$ relative to threshold $\Gamma_\star$.
    The limitations of the leading-order rate approximation are evident for large values of $-\ln{r}$.
    The indicated point corresponds to the amplifier considered in \cref{tab:vopo}.
    }
        \label{fig:finesse}
\end{figure*}

Often, the amplifier performance is characterized by the nonlinear gain, a scalar quantity found by injecting a coherent amplitude $\symbf{a}^{(2)}(0) = \dfrac{1}{\sqrt{2}}\displaystyle\begin{bmatrix} \rme^{-\rmi\Psi} \\ \rme^{+\rmi\Psi} \end{bmatrix}$ with some phase $\Psi$ into port 2 at the carrier frequency, measuring the transmission from port 1, and comparing to the same measurement with the nonlinearity turned off:
    \begin{align}
      \mathcal{G} &= \frac{\left|\underline{\symsf{G}}(0) \symbf{a}^{(2)}(0)\right|^2}{\left.\left|\underline{\symsf{G}}(0) \symbf{a}^{(2)}(0)\right|^2\right|_{\Gamma=0}} \label{eq:Gnl} \\
        &= \frac{(1 - r)^2 \left|1 - r\cosh\Gamma + r\rme^{2\rmi(\Psi - \pump{\psi})}\sinh{\Gamma}\right|^2}{(1 - 2r\cosh{\Gamma} + r^2)^2} \\
        &\xrightarrow{\Psi \rightarrow \pump{\psi}} \left(\frac{1 - r}{1 - r\rme^{+\Gamma}}\right)^2,
        \label{eq:Gnl 0}
    \end{align}
which agrees with previous expressions when $r_2 \rightarrow 1$ and $\xtal{t} \rightarrow 1$~\cite[Appx.~B]{Ganapathy:2022hgu}.
To leading order in the rates, \cref{eq:Gnl} becomes
\begin{equation}
    \mathcal{G} \simeq \frac{1 + 2x \cos{2(\Psi - \pump{\psi})} + x^2}{(1 - x^2)^2},
    \label{eq:Gnl ll}
\end{equation}
which agrees with previous expressions derived under the Collett--Gardiner formalism~\cite[Eq.~(3.116)]{McKenzieThesis}.
Since $x = -\Gamma/\ln{r}$ exactly, one can invert \cref{eq:Gnl 0} to find the relation (for $\Psi = \pump{\psi}$)
\begin{align}
    x &= 1 - \frac{\ln\left(1 - \dfrac{1-r}{\sqrt{\mathcal{G}}}\right)}{\ln r} \\
    &\simeq 1 - \frac{1}{\sqrt{\mathcal{G}}};
\end{align}
for the amplifier considered in \cref{tab:vopo}, this approximation is good to better than 1\%.

\begin{widetext}
Information is also available from measurements of the amplifier response taken at a pair of audio sideband frequencies $\pm\Omega_\text{a}$ well within the amplifier bandwidth ($\Omega_\text{a} \ll \left|s_{\text{P}-}\right|$)~\cite{Ganapathy:2022hgu}.
This method relies on the singular value decomposition of $\underline{\mathbb{G}}(0)$, which can be deduced from \cref{eq:Gsupp dc} and \cref{sec:svd}:
\begin{equation}
    \underline{\mathbb{G}}(\Omega_\text{a}) \approx \underline{\mathbb{G}}(0) =
        \begin{bmatrix*}[r] \cos{\pump{\psi}} & \sin{\pump{\psi}} \\ -\sin{\pump{\psi}} & \cos{\pump{\psi}} \end{bmatrix*}
            \begin{bmatrix} \frac{1}{1 - r\rme^{+\Gamma}} & 0 \\ 0 & \frac{1}{1 - r\rme^{-\Gamma}} \end{bmatrix}
        \begin{bmatrix*}[r] \cos{\pump{\psi}} & -\sin{\pump{\psi}} \\ \sin{\pump{\psi}} & \cos{\pump{\psi}} \end{bmatrix*},
\end{equation}
revealing the singular values $\Sigma_{\text{c},\text{s}} = \left.1\middle/\left(1 - r\rme^{\pm\Gamma}\right)\right. \simeq \left.1\middle/\fund{\tau}(\kappa \mp g)\right.$, with the upper sign referring to the cosine quadrature and the lower sign to the sine quadrature.
As described previously~\cite{Ganapathy:2022hgu}, it is then possible to deduce $\Sigma_\text{c} / \Sigma_\text{s}$, and hence $x$, from the optical beat note of the transmitted sidebands exiting port 1 against a local oscillator field at the carrier frequency.
\end{widetext}

The nonlinear gain is a dc measurement, and the audio diagnostic, while ac, is made within the amplifier's bandwidth, effectively probing its dc response only.
However, there is also information about the amplifier performance available from its ac frequency response.
For example, any measurement from which one can estimate the poles $s_{\text{P}\pm}$ (\cref{eq:sP}) allows for an estimate of $r$ and $\Gamma$ independently.
To leading order in the rates, we have $\tfrac{1}{2}(s_{\text{P}+} + s_{\text{P}-}) \simeq -\kappa$ and $\tfrac{1}{2}(s_{\text{P}+} - s_{\text{P}-}) \simeq -g$, and hence,
\begin{equation}
    \frac{s_{\text{P}+}}{s_{\text{P}-}} \simeq \frac{1 + x}{1 - x}.
\end{equation}
This approximate expression also shows that for reasonably high normalized interactions, the measurement of the frequency response must approach or exceed one decade in bandwidth to measure these poles.
Measurement of $s_{\text{P}\pm}$ is possible by measurement of $\underline{\mathbb{G}}(s)$, which can be achieved by the injection of fields from port 2 and the measurement of their beat note in transmission at port 1, similar to the audio diagnostic procedure.

If injection into port 1 is possible, using the same path taken by $\symbf{a}^{(1)}$, then by measuring in reflection of port 1 it becomes possible to extract both $x$ and the escape efficiency $\eta$ using the quadrature reflection function $\mathbb{R}(s)$.
The reflection function, like the gain function, contains the two poles $s_{\text{P}\pm}$ that can be used to extract $x$.
Then following the discussion in \cref{sec:R1}, at $\pump{\psi} = 0$ the ratio of the zero $s_{\text{Rq}++}$ in the cosine quadrature to the zero $s_{\text{Rq}+-}$ in the sine quadrature is given to leading order in the rates by
\begin{equation}
    \frac{s_{\text{Rq}++}}{s_{\text{Rq}+-}} \simeq \frac{1 - x - 2\eta}{1 + x - 2\eta},
\end{equation}
which facilitates the extraction of $\eta$.
Without appealing to the leading-order rate approximation, or to the condition that the measurement occur at $\pump{\psi} = 0$, it is still possible mathematically that from the frequency responses $\mathbb{R}_{11}(s)$ and $\mathbb{R}_{22}(s)$, which in total contain the two real poles $s_{\text{P}\pm}$ and the four real zeros $s_{\text{Rq}\pm\pm'}$ (\cref{eq:sRq quadratic,eq:sRq abc}), one can extract the six quantities $r$, $\Gamma$, $r_1$, $t_1$, $\pump{\psi}$, and $\fund{\tau}$, and thereby jointly determine $x$ and $\eta$.

\section{Variances and squeezing}
\label{sec:noise}

We now examine the statistics of the output of an optical parametric amplifier when all ingoing fundamental fields are initialized at all Fourier frequencies with unsqueezed vacuum, notwithstanding a few discrete frequencies that may have coherent amplitudes for control purposes.
This is the usual operating mode for amplifiers used to produce broadband squeezed vacuum, and it is also a key ingredient for estimating the signal-to-noise ratio when amplifying coherent fields.
We will compute the noise of this squeezed vacuum and examine the contributions to it, both using the full expressions for the cavity dynamics and the leading-order Padé expansions.
The general strategy for computing vacuum variances from the input--output relations follows the development given in Ref.~\cite[\S{}5]{Danilishin:2012fa}.

\begin{figure*}
    \centering
    \includegraphics[width=\textwidth]{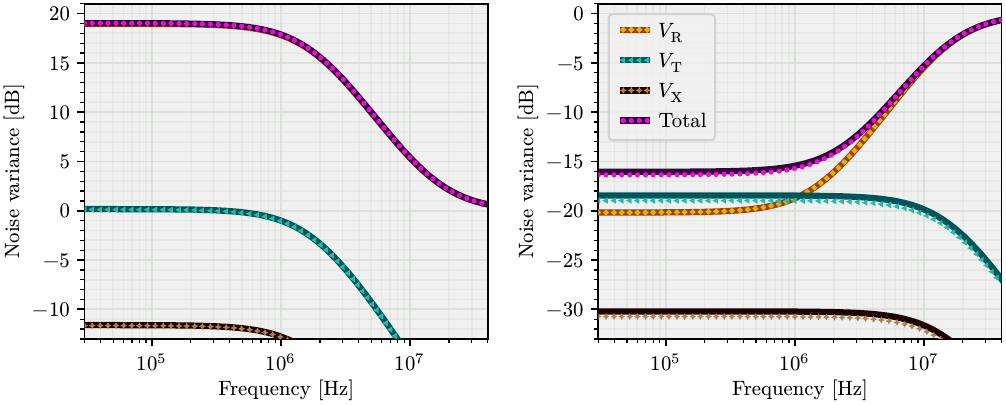}
    \caption{Noise variances leaving mirror~1, relative to shot noise. \emph{Left:} cosine quadrature; \emph{right:} sine quadrature; here the quadratures are plotted at a pump phase $\pump{\psi} = 0$.
    Solid lines are the exact computation; dotted lines are the zero--pole--gain Padé expansions, approximated to leading order in the cavity rates.
    The contributions are $V_\text{R}$, the variance from the field $\mathbb{a}^{(1)}$ impinging on mirror~1; $V_{\text{T}}$, the variance from the field $\mathbb{a}^{(2)}$ impinging on mirror~2; and $V_{\text{X}}$, the variance from the field $\mathbb{a}^{(\text{X})}$ entering via the crystal loss; all fields are initialized in the vacuum state.
    Refer to \cref{sec:noise} for formulae.}
    \label{fig:opa io V1}
\end{figure*}

From the quadrature equivalent of the sideband relation \cref{eq:a1prime}, the quadrature field vector $\mathbb{a}^{(1')}$ exiting the external side of mirror~1 is the superposition of several fields: first, the field $\mathbb{a}^{(1)}$ impinging externally from mirror~1 and becoming squeezed by the reflection $\mathbb{R}$ of the cavity (\cref{sec:R1}); second, the field $\mathbb{a}^{(2)}$ entering mirror~2 and being transmitted to mirror~1 by $\mathbb{T}$; third, the field $\mathbb{a}^{(\text{X})}$ entering via the crystal loss and similarly being transmitted to mirror~1 via $\mathbb{T}^{(\text{X})}$:
\begin{equation}
    \mathbb{a}^{(1')}(s) = \mathbb{R}(s)\,\mathbb{a}^{(1)}(s) + \mathbb{T}(s)\,\mathbb{a}^{(2)}(s) + \mathbb{T}^{(\text{X})}(s)\,\mathbb{a}^{(\text{X})}(s),
    \label{eq:a1prime q}
\end{equation}
where, comparing with \cref{eq:a1prime}, we have absorbed the various propagation matrices $\left\{\mathbb{L}_{j}\right\}$ (the quadrature equivalents of \cref{eq:Lprop}) into the fields $\left\{\mathbb{a}^{(j)}\right\}$ for $j = 1, 2, \text{X}$, which only has the effect of rotating the quadrature amplitudes.

With the exception of isolated frequencies used for coherent control, the input fields are assumed to be in the vacuum state (including at the carrier frequency $s = 0$), which means their cross spectral density matrices are $\left\langle \mathbb{a}^{(j)} {\mathbb{a}^{(j)}}^\dagger \right\rangle = \mathbb{1}$ in normalized units for $\mathbb{a}^{(j)} \in \left\{\mathbb{a}^{(1)}, \mathbb{a}^{(2)}, \mathbb{a}^{(\text{X})}\right\}$.
We can then compute the cross-spectral density matrix of the field quadratures exiting mirror~1:
\begin{equation}
    V^{(1'1')}(\Omega) = \left\langle\mathbb{a}^{(1')}{\mathbb{a}^{(1')}}^\dagger\right\rangle
    = \mathbb{R} \mathbb{R}^\dagger + \mathbb{T} \mathbb{T}^\dagger + \mathbb{T}^{(\text{X})} {\mathbb{T}^{(\text{X})}}^\dagger.
\end{equation}
We plot the elements of this matrix in \cref{fig:opa io V1}, using the exact form of the quadrature matrices and also the Padé-expanded forms assuming low transmissivity.
For the parameters that we have assumed, the noise variance in the squeezed quadrature below \qty{1}{\MHz} is dominated by two nearly commensurate contributions from the vacuum entering via mirror~2 and the vacuum impinging on mirror~1.
Above \qty{1}{\MHz}, the phase-sensitive deamplification of the vacuum state impinging on mirror~1 is reduced because the relevant sidebands are not fully within the amplifier's bandwidth, as discussed in \cref{sec:R1}; the total noise variance is therefore dominated by the vacuum state impinging on mirror~1.
Having assumed a crystal loss $\xtal{r}^2 \approx \qty{100}{ppm}$, the noise due to the vacuum entering via this loss channel is subdominant at all frequencies.
A tenfold higher crystal loss would, however, contribute a noise in the squeezed quadrature that competes with the noises entering at mirrors~1 and~2.

Now we investigate the mathematical form of the squeezed vacuum level at dc at a pump phase $\pump{\psi} = 0$.
We note the following limits for reflection:
\begin{align}
    \mathbb{R}(0) &\xrightarrow{r_1^2 + t_1^2 \rightarrow 1} \begin{bmatrix}
        \frac{r_1^2 - r\rme^{+\Gamma}}{r_1(1 - r \rme^{+\Gamma})} &
            0 \\
            0 &
            \frac{r_1^2 - r \rme^{-\Gamma}}{r_1(1 - r \rme^{-\Gamma})} \\
    \end{bmatrix} \\
    &\xrightarrow{\Gamma \rightarrow \Gamma_\star} \begin{bmatrix}
        \infty & 0 \\
        0 & \frac{r_1^2 - r^2}{r_1(1 - r^2)} \\
    \end{bmatrix}.
\end{align}
which agrees with \textcite[Eq.~(2.55)]{GanapathyThesis}.
Similarly,
\begin{equation}
    \mathbb{T}(0) = \begin{bmatrix}
        \frac{t_1 t_2}{1 - r \rme^{+\Gamma}} &
            0 \\
            0 &
            \frac{t_1 t_2}{1 - r \rme^{-\Gamma}} \\
    \end{bmatrix} \xrightarrow{\Gamma \rightarrow \Gamma_\star} \begin{bmatrix}
        \infty & 0 \\
        0 & \frac{t_1 t_2}{1 - r^2} \\
    \end{bmatrix}
\end{equation}
and
\begin{equation}
    \mathbb{T}^{(\text{X})}(0) = \begin{bmatrix}
        \frac{t_1 r_2 \xtal{r}}{1 - r \rme^{+\Gamma}} &
            0 \\
            0 &
            \frac{t_1 r_2 \xtal{r}}{1 - r \rme^{-\Gamma}} \\
    \end{bmatrix} \xrightarrow{\Gamma \rightarrow \Gamma_\star} \begin{bmatrix}
        \infty & 0 \\
        0 & \frac{t_1 r_2 \xtal{r}}{1 - r^2} \\
    \end{bmatrix},
\end{equation}
so that the squeeze level at dc is
\begin{subequations}
\begin{align}
    V^{(1'1')}_{\text{s}}(0) &= \frac{\left(r_1 - r\rme^{-\Gamma}/r_1\right)^2 + (t_1 t_2)^2 + \left(t_1 r_2 \xtal{r}\right)^2}{\left(1 - r\rme^{-\Gamma}\right)^2} \\
    &\xrightarrow{\Gamma \rightarrow \Gamma_\star} \frac{\left(r_1 - r^2 / r_1\right)^2 + (t_1 t_2)^2 + (t_1 r_2 \xtal{r})^2}{(1 - r^2)^2},
    \label{eq:V11 max}
\end{align}
\end{subequations}
where (again) we assumed $r_1^2 + t_1^2 = 1$.

\begin{widetext}
We can make a leading-order approximation to \cref{eq:a1prime q} by Taylor expanding the numerators and denominators of $\mathbb{R}(0)$, $\mathbb{T}(0)$, and $\mathbb{T}^{(\text{X})}(0)$, which yields the following relations between quadratures:
\begin{equation}
    a^{(1')}_{\text{c,s}}(0) = \frac{(\kappa - 2\kappa_1 \mp g) a^{(1)}_{\text{c,s}}(0) + 2\sqrt{\kappa_1 \kappa_2} a^{(2)}_{\text{c,s}}(0) + 2\sqrt{\kappa_1 \kappa_{\text{X}}} a^{(\text{X})}_{\text{c,s}}(0)}{\kappa \mp g}
\end{equation}
where the upper signs refer to the cosine quadrature and the lower signs to the sine quadrature with $\pump{\psi} = 0$.
This is identical to any number of expressions found in the gravitational-wave literature derived under the Collett--Gardiner formalism (e.g., the dc expression of Ref.~\cite[Eq.~(4.58)]{MansellThesis}), keeping in mind our sign convention for fields at reflective interfaces.
This leads to the standard expressions for the dc quadrature variance in the leading-order rate approximation:
\begin{equation}
    V^{(1'1')}_{\text{c},\text{s}}(0) \simeq 1 \pm \frac{4\kappa_1 g}{(\kappa \mp g)^2} = 1 \pm \frac{4\eta x}{(1 \mp x)^2},
    \label{eq:V11 max ll}
\end{equation}
which exploits the rate, gain, and escape efficiency relations in \cref{sec:rates}.
\end{widetext}

\begin{figure}
    \centering
    \includegraphics[width=\columnwidth]{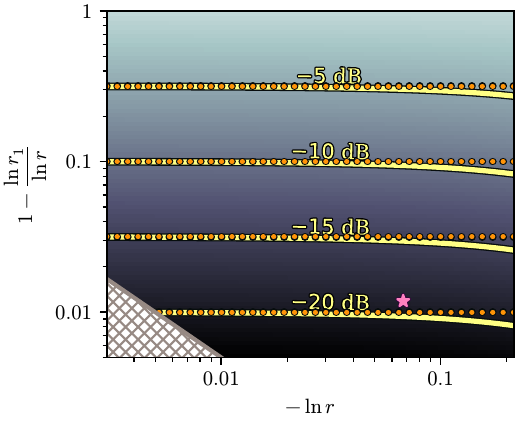}
    \caption{Maximum attainable squeezing as a function of $-\ln{r} = \kappa\fund{\tau}$ and $-\ln{r_1} = \kappa_1\fund{\tau}$, with $-\ln{r_2} = \kappa_2\fund{\tau}$ allowed to vary.
    The horizontal axis is the total decay rate normalized by the cavity free spectral range; the vertical axis is the escape loss $1 - \eta$.
    Solid lines show the exact expression from \cref{eq:V11 max}; dotted lines show the approximation from \cref{eq:V11 max ll} with $x \rightarrow 1$.
    The hatched region is forbidden by a fixed amount of intracavity crystal loss $\xtal{r}^2 = 1 - \rme^{-\mathcal{A}\xtal{l}}$, with values of $\mathcal{A}$ and $\xtal{l}$ taken from \cref{tab:vopo}.
    The indicated point corresponds to the maximal squeezing attainable by the amplifier in \cref{tab:vopo} (with $\Gamma\rightarrow\Gamma_{\star}$).
    Evidently, for given values of $-\ln{r_1}$ and $-\ln{r}$, the use of the approximate formulae will slightly overpredict the maximum amount of attainable squeezing, especially at large decay rates.
    }
    \label{fig:max squeezing}
\end{figure}

In \cref{fig:max squeezing} we compare the exact and approximate expressions for the maximum attainable squeezing at dc.
As expected, the approximations are increasingly exact as $-\ln{r} = \kappa\fund{\tau}$ becomes small.
For the amplifier discussed in this work, the approximated contributions to the overall squeezed variance are accurate to about \qty{0.5}{\dB} or better (right side of \cref{fig:opa io V1}).
Although we have not done so here, we note again that in the presence of fluctuations of the quadrature angle of the readout (or coherent signal) relative to the amplifier, we could estimate the effective noise variance in the usual fashion: namely, if the fluctuations are characterized by an rms phase $\theta \ll 1$, the effective variance of the squeezed vacuum (nominally in the sine quadrature) is approximately $V_{\text{s}}^{(1'1')} \cos^2 {\theta} + V_{\text{c}}^{(1'1')}\sin^2{\theta}$~\cite{Aoki:2005gqw}.
Similarly, simple losses in the readout chain can be modeled with additional beamsplitter operations that superpose $\mathbb{a}^{(1')}$ with additional unsqueezed vacuum fields $\left\{\mathbb{a}^{(\text{vac},j)}\right\}$.

\section{Conclusion}
\label{sec:conclusion}

We have derived input--output relations for a resonant optical parametric amplifier, starting directly from exact relations between the optical fields entering, circulating in, and exiting the resonant cavity.
To leading order in the decay and gain rates, these relations are consistent with previous Hamiltonian-based analyses, although our relations include some high-frequency effects from the finite free spectral range of the amplifier cavity, which are not captured in a Hamiltonian analysis.
We have used our relations to produce frequency-dependent estimates of the quadrature variance of the squeezed vacuum state produced by such an amplifier.
Overall, the leading-order rate approximations to the exact relations are quite accurate in both their amplitude and frequency dependence; the discrepancy in the contributions to the quadrature variance is less than \qty{0.5}{\dB} for the amplifier we considered.
The discrepancy may be larger in amplifiers with larger total decay rate.
Along the way, we have emphasized that the input--output relations encode information about the amplifier performance that is accessible by frequency-response measurements, complementary to dc measurements such as the nonlinear gain.

There are several theoretical extensions to the work presented here, perhaps the most immediate being the addition of detuning to the cavity gain and reflection functions, which is relevant for cavity control.
Additionally, we have not attempted to derive expressions for quadrature phase noise from mechanisms in the amplifier that rotate the squeeze angle.
Also, our formalism could be extended to track the evolution of the pump modes, as has been done in previous Hamiltonian-based analyses (e.g., \textcite{Goda:2005ytq}).
Finally, the effects of transverse spatial modes will complicate the analysis, at minimum because of additional effects within the nonlinear medium that alter the coupling strength and phase matching~\cite{1968JAP....39.3597B,2007OExpr..15.7211L,BoydNonlinear}.
A description of parametric amplification in multiple spatial modes simultaneously is sometimes desired~\cite{Heinze:2022vky} and could be described by extending this work to higher dimension.

\begin{acknowledgments}
The author thanks Kevin Kuns, James Gardner, Aaron Markowitz, Francisco Salces-Cárcoba, and Sander Vermeulen for thoughtful comments on this work.
The symbolic computations in this work were aided by \texttt{sympy}~\cite{Sympy}.
The author gratefully acknowledges the support of the United States National Science Foundation (NSF) for the construction and operation of the LIGO Laboratory and Advanced LIGO as well as the Science and Technology Facilities Council (STFC) of the United Kingdom, and the Max-Planck-Society (MPS) for support of the construction of Advanced LIGO.
Additional support for Advanced LIGO was provided by the Australian Research Council (ARC).
LIGO was constructed by the California Institute of Technology and Massachusetts Institute of Technology with funding from the National Science Foundation and operates under Cooperative Agreement PHY--18671764464, which also funds the author's work.
Advanced LIGO was built under NSF PHY--18680823459. The A+ upgrade to Advanced LIGO is supported by NSF PHY--1834382 and STFC ST/S00246/1, with additional support from the ARC.
\end{acknowledgments}

\appendix

\section{Sidebands and quadratures}
\label{sec:sidebands}

This appendix collects certain conventions for handling sideband and quadrature representations of modulated optical fields.
Sideband fields and quadrature fields are related by
\begin{equation}
    a_{\text{c}} = \frac{a_+ + a_-^\dagger}{\sqrt{2}} \quad \text{and} \quad a_{\text{s}} = \frac{a_+ - a_-^\dagger}{\sqrt{2}\rmi},
\end{equation}
which is equally valid in the time or frequency domain.
We relate the time and frequency domains via the bilateral Laplace transform
\begin{equation}
    a_\pm(s) = \int\limits_{\mathclap{-\infty}}^{\mathclap{+\infty}} \rmd{t}\,\rme^{-st}\,a_\pm(t),
\end{equation}
where $s$ is complex-valued, and angular Fourier frequencies $\Omega$ are found on the imaginary line $s = \rmi\Omega$.
Specifically in the time domain, a modulated field at carrier frequency $\omega_0$ and with carrier amplitude $A_0$ can be written
\begin{equation}
    A(t) = \tfrac{1}{2} \left(A_0 + a_+(t) \rme^{+\rmi\Omega t} + a_-(t) \rme^{-\rmi\Omega t}\right) \rme^{+\rmi\omega_0 t} + \text{cc}.
\end{equation}
Therefore, introducing a time delay $\tau$ and comparing $A(t-\tau)$ to $A(t)$ enables us to identify the convention for the time-delay operation acting on $s$-domain sideband amplitudes:
\begin{equation}
    \left.\begin{bmatrix} a_+(s) \\ a_-^\dagger(s) \end{bmatrix}\right|_{t-\tau}
        = \rme^{-\rmi\Omega\tau} \begin{bmatrix} \rme^{-\rmi\omega_0\tau} & 0 \\ 0 & \rme^{+\rmi\omega_0\tau} \end{bmatrix}
        \left.\begin{bmatrix} a_+(s) \\ a_-^\dagger(s) \end{bmatrix}\right|_{t},
    \label{eq:sb td}
\end{equation}
so long as we can assume that the amplitudes $a_\pm$ vary more slowly in time than $1/\Omega$.

\section{Solution of the parametric interaction equation}
\label{sec:su11}

This section solves the evolution of upper and lower plane wave sidebands, traveling in the $+z$ direction, coupled to each other via a nonlinear medium extending from $z=0$ to $+\xtal{l}$, which is described by \cref{eq:opa diff eq}.
We define
\begin{equation}
  \symbf{\tilde{a}}(z) = \begin{bmatrix} \rme^{+\rmi\Delta k\,z/2} & 0 \\ 0 & \rme^{-\rmi\Delta k\,z/2} \end{bmatrix} \symbf{a}(z),
  \label{eq:atilde}
\end{equation}
which upon substitution into \cref{eq:opa diff eq} results in an ordinary differential equation (ODE) with constant coefficients:
\begin{equation}
  \frac{\rmd{\symbf{\tilde{a}}}}{\rmd{z}} =
    \frac{1}{2}\begin{bmatrix}
      +\rmi\Delta k & -\rmi\gamma \\
      +\rmi\gamma^* & -\rmi\Delta k
    \end{bmatrix}
    \symbf{\tilde{a}}(z).
   \label{eq:opa diff eq 1}
\end{equation}
By defining the matrices
\begin{equation}
  \symsf{1} = \begin{bmatrix*}[r] 1 & 0 \\ 0 & 1 \end{bmatrix*} \quad
  \symsf{K}_x = \begin{bmatrix*}[r] 0 & 1 \\ 1 & 0 \end{bmatrix*} \quad
  \symsf{K}_y = \begin{bmatrix*}[r] 0 & \rmi \\ -\rmi & 0 \end{bmatrix*} \quad
  \symsf{K}_z = \begin{bmatrix*}[r] \rmi & 0 \\ 0 & -\rmi \end{bmatrix*}
  \label{eq:su11}
\end{equation}
we can rewrite \cref{eq:opa diff eq 1} as
\begin{equation}
  \frac{\rmd\symbf{\tilde{a}}}{\rmd{z}} = \left(\frac{\mu}{2}\,\vec{m} \cdot \vec{\symsf{K}}\right)\symbf{\tilde{a}}(z)
  \label{eq:opa diff eq 2}
\end{equation}
where $\vec{m}\cdot\vec{\symsf{K}}$ denotes a linear combination of the three matrices $\symsf{K}_x$, $\symsf{K}_y$, and $\symsf{K}_z$ with the scalar coefficients collected into the vector
\begin{equation}
  \vec{m} = \frac{1}{\mu} \left(
     |\gamma|\cos{2\pump{\psi}}, \;
    -|\gamma|\sin{2\pump{\psi}}, \;
    \Delta k
    \right)
\end{equation}
with $\mu = \sqrt{|\gamma|^2 - (\Delta k)^2}$.
Then by introducing the integrating factor $\rme^{-(\mu/2)\left(\vec{m}\cdot\vec{\symsf{K}}\right)z}$, and noting that any matrix $\symsf{B}$ commutes with $\rme^{\symsf{B}z}$ for all $z$, we can integrate both sides from $z=0$ to $+\xtal{l}$ to find
\begin{equation}
  \symbf{\tilde{a}}(\xtal{l}) = %
  \exp\left(\frac{\mu\xtal{l}}{2}\left(\vec{m}\cdot\vec{\symsf{K}}\right)\right)\symbf{\tilde{a}}(0).
  \label{eq:opa diff eq 3}
\end{equation}
Since $\left(\vec{m}\cdot\vec{\symsf{K}}\right)^2 = \symsf{1}$, the exponent resolves to
\begin{equation}
  \exp\left(\frac{\mu\xtal{l}}{2}\left(\vec{m}\cdot\vec{\symsf{K}}\right)\right)
    = \cosh\left(\frac{\mu\xtal{l}}{2}\right)\symsf{1}
      + \sinh\left(\frac{\mu\xtal{l}}{2}\right)\left(\vec{m}\cdot\vec{\symsf{K}}\right).
  \label{eq:exp map}
\end{equation}
Together with \cref{eq:atilde}, this gives the functional form of $\xtal{\symsf{H}}$ in \cref{eq:HX} and hence \cref{eq:HX hyperbolic}.
The two upper elements of $\xtal{\symsf{H}} - \symsf{1}$ are plotted as a function of the phase match $\Delta k\,\xtal{l}/2$ in \cref{fig:HX}.

One can see by inspection of \cref{eq:HX} or \cref{eq:HX hyperbolic} that $\xtal{\symsf{H}}$ is unitary, and more specifically, that $\xtal{\symsf{H}}$ belongs to $\mathrm{SU}(1,1)$.
This also follows from the fact that the matrices $\symsf{K}_x,\symsf{K}_y,\symsf{K}_z$ are representations of the elements in the Lie algebra $\mathfrak{su}(1,1)$ that generate $\mathrm{SU}(1,1)$ by the exponentiation in \cref{eq:opa diff eq 3}~\cite{Yurke:1986ipz,Chiribella:2006vet}.

To arrive at a closed form for $\xtal{\symsf{H}}$, we relied on the ability to transform \cref{eq:opa diff eq}, which is an ODE of the form $\rmd{\symbf{a}}/\rmd{z} = \symsf{Q}(z) \symbf{a}(z)$, into an ODE with constant coefficients by change of variable in \cref{eq:atilde}.
As remarked in the main text, various effects, particularly phase shifts from focusing, preclude this kind of transformation in general.
However, so long as $\symsf{Q}(z)$ belongs to $\mathfrak{su}(1,1)$, the infinitesimal transformation $\symbf{a}(z+\rmd{z}) = \rme^{\symsf{Q}(z)\rmd{z}}\symbf{a}(z)$ likewise belongs to $\mathrm{SU}(1,1)$, and thus so will the overall product of transformations from $z=0$ to $+\xtal{l}$.
This underlies the assertion in the main text that even in this more complicated situation $\xtal{\symsf{H}}$ must still be able to be parametrized in a form appropriate for $\mathrm{SU}(1,1)$, e.g.,
\begin{equation}
  \begin{bmatrix}
    \rme^{-\rmi\nu} \cosh{w} & \rme^{-\rmi\rho}\sinh{w} \\
    \rme^{+\rmi\rho}\sinh{w} & \rme^{+\rmi\nu} \cosh{w}
  \end{bmatrix}
\end{equation}
for a nonnegative parameter $w$ and two real parameters $\nu$ and $\rho$, the three of which may, however, depend nontrivially on the structure and properties of the nonlinear medium.

\begin{widetext}
Next, we examine an approximate form of $\xtal{\symsf{H}}$.
By taking $\mu \simeq \rmi\Delta k$, we have $\sinh\Gamma \simeq (|\gamma|\xtal{l}/2) \sinc(\Delta k \xtal{l}/2)$, where $\sinc{u} \equiv (\sin{u})/u$, and $\zeta \simeq -\Delta k\,\xtal{l}/2$. %
Then one finds, at $\pump{\psi} = 0$,
\begin{equation}
    \label{eq:HX McKenzie}
    \xtal{\symsf{H}} \xrightarrow{\mu \simeq \rmi\Delta k}
    \begin{bmatrix}
        1 & -\rmi \frac{\gamma \xtal{l}}{2} \rme^{-\rmi\Delta k\,\xtal{l}/2} \sinc{\frac{\Delta k\,\xtal{l}}{2}}\\
        +\rmi \frac{\gamma^* \xtal{l}}{2} \rme^{+\rmi\Delta k\,\xtal{l}/2} \sinc{\frac{\Delta k\,\xtal{l}}{2}} & 1
    \end{bmatrix}.
\end{equation}
This equation is consistent with some previous expressions used in the gravitational-wave literature~\cite[Eq.~(3.101)]{McKenzieThesis} \cite[Eq.~(28)]{Goda:2005ytq}, although we note that for $\gamma \ne 0$ the matrix \cref{eq:HX McKenzie} is not unitary.
In the main text we have kept the exact expression \cref{eq:HX}, particularly in the $\Delta k \rightarrow 0$ limit given by \cref{eq:HX Ganapathy}.

\begin{figure*}
    \centering
    \includegraphics[width=\textwidth]{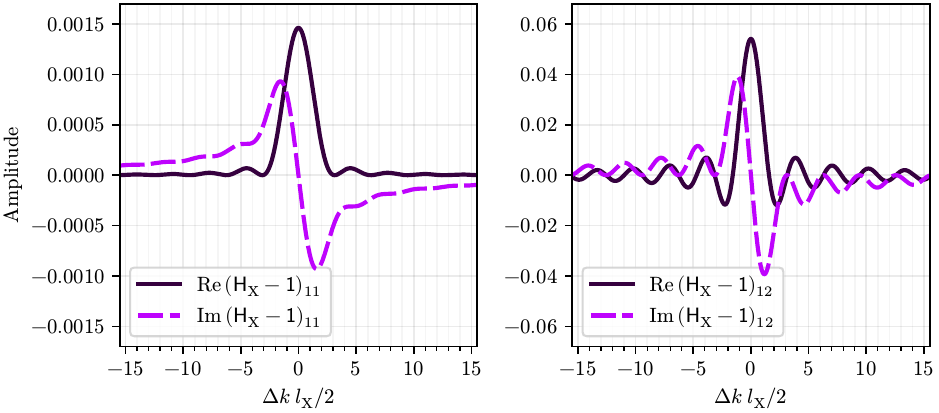}
    \caption{The exact form (\cref{eq:HX}) of the single-pass nonlinear coupling matrix $\xtal{\symsf{H}}$ for the interaction of plane waves, plotted here with the identity $\symsf{1}$ subtracted, as a function of the phase mismatch $\Delta k\,\xtal{l}/2$.
    Only the two upper elements $(\xtal{\symsf{H}} - \symsf{1})_{11}$ and $(\xtal{\symsf{H}} - \symsf{1})_{12}$ are plotted, as the lower elements are the complex conjugates of the upper elements.
    The element $(\xtal{\symsf{H}})_{11}$ indicates the evolution of the complex amplitude $a_+$ of the upper sideband during propagation through the nonlinear crystal; the element $(\xtal{\symsf{H}})_{12}$ indicates how the nonlinear interaction generates a lower sideband with conjugate complex amplitude $a_-^\dagger$ from the upper sideband $a_+$ in the presence of a pump field.
    Here we took values from \cref{tab:vopo}.
    Note that the vertical scales of the plots are different.}
    \label{fig:HX}
\end{figure*}

To include the effect of a nonzero power loss $\mathcal{A} > 0$ per unit length (from effects such as absorption or scatter), one adds $-\mathcal{A}/2$ to the diagonals of \cref{eq:opa diff eq}, which leads to the replacement $\frac{\mu}{2}\left(\vec{m}\cdot\vec{\symsf{K}}\right) \rightarrow -\frac{\mathcal{A}}{2}\symsf{1} + \frac{\mu}{2}\left(\vec{m}\cdot\vec{\symsf{K}}\right)$ in \cref{eq:opa diff eq 2}.
The exponentiation $\rme^{-\frac{\mathcal{A}}{2}\symsf{1} + \frac{\mu}{2}\left(\vec{m}\cdot\vec{\symsf{K}}\right)}$ factorizes to $\rme^{-\mathcal{A}\xtal{l}/2}\rme^{(\mu\xtal{l}/2)\left(\vec{m}\cdot\vec{\symsf{K}}\right)}$ because $\symsf{1}$ commutes trivially with $\symsf{K}_j$ for $j \in \{x,y,z\}$.
This justifies the claim in \cref{sec:absorption} that the lossy evolution of $\symbf{a}(0)$ into $\symbf{a}(+\xtal{l})$ is the same as the lossless case, except for the appearance of an amplitude prefactor $\rme^{-\mathcal{A}\xtal{l}/2}$.
The coupling of a secondary mode $\symbf{a}^\text{(X)}$ to $\symbf{a}$ with an overall amplitude $\sqrt{1 - \rme^{-\mathcal{A}\xtal{l}}}$ preserves the unitarity of the lossy interaction.

\section{Comparison with Collett--Gardiner formalism}
\label{sec:cg}

This section gives the first steps in how the formalism presented in this work can be brought into contact with the Hamiltonian analysis of \textcite{Collett:1984ulf}.
Specifically, we want to compare our input--output relation $\underline{\symsf{G}}$ derived in \cref{sec:Gsupp} with the quantum Langevin equations that emerge from the Heisenberg equations of motion for the quantum field operators (e.g., in Refs.~\cite{McKenzieThesis,Goda:2005ytq}).
To do this, we transform our \cref{eq:a0} into the time domain and then make the mean-field approximation, averaging the cavity dynamics over the round-trip time.

Specifically, we write the total round-trip phase delay for the fundamental field as $\fund{\omega} \fund{\tau} = \fund{\omega} \fund{\tilde{\tau}} + \fund{\Phi}$, with $\fund{\omega}\fund{\tilde{\tau}}$ being an integer multiple of $2\pi$, and $\fund{\Phi}$ being the remaining phase delay.
Then we have
\begin{align}
    \dot{\symbf{a}}(t) &\triangleq \frac{\symbf{a}(t) - \symbf{a}(t - \fund{\tilde{\tau}})}{\fund{\tilde{\tau}}} \\
    &= \frac{t_1}{\fund{\tilde{\tau}}} \symbf{a}^{(1)}(t)
    - \frac{r_1 t_2}{\fund{\tilde{\tau}}} \symbf{a}^{(2)}(t - \tau_{12})
    + \frac{r\rme^{-\rmi\fund{\symbfsf{\Phi}}}\xtal{\symsf{H}} - \symsf{1}}{\fund{\tilde{\tau}}} \symbf{a}(t) + \frac{r_1 r_2 \xtal{r}}{\fund{\tilde{\tau}}} \symbf{a}^{(\text{X})}(t - \tau_{12} - \tau_{1\text{X}} - \xtal{\tau}),
\end{align}
where to get the second line we used \cref{eq:sb td} to find the time-domain equivalent of \cref{eq:a0}.
This procedure also assumes the signal and idler frequencies $\fund{\omega} \pm \Omega \simeq \fund{\omega}$.
The quantity $r\rme^{-\rmi\fund{\symbfsf{\Phi}}}\xtal{\symsf{H}}$, with $\rme^{-\rmi\fund{\symbfsf{\Phi}}} = \begin{bmatrix} \rme^{-\rmi\fund{\Phi}} & 0 \\ 0 & \rme^{+\rmi\fund{\Phi}} \end{bmatrix}$, is the time-domain analogue of $\symsf{G}$ (\cref{eq:G}).
Now we take the following approximations for small rates and detunings:
\begin{align}
    r_1 r_2 \xtal{t} = r &\rightarrow 1 - \kappa\fund{\tilde{\tau}} \\
    \rme^{\pm\rmi\fund{\Phi}} &\rightarrow 1 \pm \rmi\fund{\Phi} \\
    \cosh\Gamma &\rightarrow 1 \\
    \sinh\Gamma &\rightarrow g\fund{\tilde{\tau}}
\end{align}
and we drop any terms that involve products of the losses, phases, or rates.
Then, using our definition of $\xtal{\symsf{H}}$ (\cref{eq:HX}) with perfect phase matching, we have
\begin{equation}
    \dot{\symbf{a}}(t) = \sqrt{\frac{2\kappa_1}{\fund{\tilde{\tau}}}} \symbf{a}^{(1)}(t) - \sqrt{\frac{2\kappa_2}{\fund{\tilde{\tau}}}} \symbf{a}^{(2)}(t - \tau_{12}) +  \begin{bmatrix} -\kappa - \left.\rmi\fund{\Phi}\middle/\fund{\tilde{\tau}}\right. & g \rme^{-2\rmi\pump{\psi}} \\ g\rme^{+2\rmi\pump{\psi}} & -\kappa + \left.\rmi\fund{\Phi}\middle/\fund{\tilde{\tau}}\right. \end{bmatrix} \symbf{a}(t)
    + \sqrt{\frac{2\xtal{\kappa}}{\fund{\tilde{\tau}}}} \symbf{a}^{(\text{X})}(t - \tau_{12} - \tau_{1\text{X}} - \tau_{2\text{X}}).
    \label{eq:adot}
\end{equation}
\end{widetext}

Comparing with the definitions in \textcite[\S~3.5.2]{McKenzieThesis}, we see that we should make the following identifications for the rates and detunings:
{\allowdisplaybreaks
\begin{subequations}
\begin{align}
    \fund{\kappa}^{\text{(out)}} &= \kappa_1 \\
    \fund{\kappa}^{\text{(in)}} &= \kappa_2 \\
    \fund{\kappa}^{\text{(loss)}} &= \xtal{\kappa} = \frac{\mathcal{A}\xtal{l}}{2\fund\tau} \\
    \fund{\kappa} &= \fund{\kappa}^{\text{(in)}} + \fund{\kappa}^{\text{(out)}} + \fund{\kappa}^{\text{(loss)}} \\
    \fund{\Delta} &= \left.\fund{\Phi} \middle/ \fund{\tilde{\tau}}\right. \\
    q &= g\rme^{-2\rmi\pump{\psi}}
  \end{align}
and for the field quantities
\begin{align}
    \symbf{a}^{\text{(out)}} &= \left.\symbf{a}^{(1)} \middle/ \sqrt{\fund{\tilde{\tau}}} \right.{} \\
    \symbf{a}^{\text{(in)}} &= \left.\symbf{a}^{(2)} \middle/ \sqrt{\fund{\tilde{\tau}}} \right.{} \\
    \symbf{a}^{\text{(loss)}} &= \left.\symbf{a}^{(\text{X})} \middle/ \sqrt{\fund{\tilde{\tau}}} \right.{},
\end{align}
\end{subequations}}
at which point the matrix $\dfrac{1}{\fund{\tilde{\tau}}} (r\rme^{-\rmi\fund{\symbfsf{\Phi}}}\xtal{\symsf{H}} - \symsf{1})$ becomes $\symsf{M}_{\symbf{a}}$ as defined by \textcite[Eq.~(3.113)]{McKenzieThesis}.
By a Laplace transform of \cref{eq:adot}, we should therefore conclude that under the Collett--Gardiner model, the internal gain is $\left((1 + s\tilde{\tau}) \symsf{1} - r\rme^{-\rmi\fund{\symbfsf{\Phi}}}\xtal{\symsf{H}}\right)^{-1}$.
Computing this expression indeed reproduces our leading-order approximation for $\underline{\symsf{G}}$ (\cref{eq:Gsupp zpk,eq:Gsupp zpk ll dc,eq:sT ll}), except the zeros at $2/\fund{\tilde{\tau}}$, associated with the cavity's finite free spectral range, are are absent.

\section{Padé expansion of closed loop gain}
\label{sec:cloop}

An exact expression for the closed-loop gain $\overline{\symsf{G}}(s)$ was given in \cref{eq:Gc}, and we Padé approximate it here and show that to leading order in the rates, it coincides with the leading-order approximation to $\underline{\symsf{G}}(s)$.

The Padé expansion of $\overline{\symsf{G}}$ is
\begin{equation}
    \overline{\symsf{G}}(s) \xrightarrow{\text{Padé}} \frac{\overline{\symsf{G}}(0) \times (1 - s\fund{\tau}/2)}{\left(1 - s/s_{\text{P}+}\right)\left(1 - s/s_{\text{P}-}\right)}
    \odot \begin{bmatrix} 1 - s/s_{\text{T}'} & 1 + s\fund{\tau}/2 \\ 1 + s\fund{\tau}/2 & 1 - s/s_{\text{T}'} \end{bmatrix}.
\end{equation}
The dc value of $\overline{\symsf{G}}$ is
\begin{align}
    \overline{\symsf{G}}(0) &= \frac{r\mathcal{F}_\Gamma^2}{\pi^2}
    \begin{bmatrix}
        \cosh{\Gamma} - r &
        \rme^{-2\rmi\pump{\psi}}\sinh{\Gamma} \\
        \rme^{+2\rmi\pump{\psi}}\sinh{\Gamma} &
        \cosh{\Gamma} - r \\
    \end{bmatrix} \\
    &\simeq \frac{1/\fund{\tau}}{\kappa^2 - g^2} \begin{bmatrix} \kappa & g \rme^{-2\rmi\pump{\psi}} \\ g \rme^{+2\rmi\pump{\psi}} & \kappa \end{bmatrix}
\end{align}
and we define the zero
\begin{align}
    s_{\text{T}'} &= \frac{2}{\fund{\tau}} \times \frac{r - \cosh{\Gamma}}{r + \cosh{\Gamma}} \\
    &\simeq -\kappa.
\end{align}

\begin{widetext}
We can transform into the quadrature basis by writing
\begin{align}
    \overline{\mathbb{G}}(s) &= \symsf{A}\?\overline{\symsf{G}}(s)\?\symsf{A}^\dagger \\
    &= \frac{r\rme^{-s\fund{\tau}}}{1 - 2r\rme^{-s\fund{\tau}}\cosh{\Gamma} + r^2 \rme^{-2s\fund{\tau}}} \nonumber \\
    &\hphantom{=} \qquad\qquad \times \begin{bmatrix}
        \cosh{\Gamma} + \cos{2\pump{\psi}} \sinh{\Gamma} -r\rme^{-s\fund{\tau}} &
        -\sin{2 \pump{\psi}} \sinh{\Gamma} \\
        -\sin{2 \pump{\psi}} \sinh{\Gamma} &
        \cosh{\Gamma} - \cos{2\pump{\psi}}\sinh{\Gamma} -r\rme^{-s\fund{\tau}} \\
    \end{bmatrix},
\end{align}
and then Padé expand as before:
\begin{equation}
    \overline{\mathbb{G}}(s) \xrightarrow{\text{Padé}} \frac{\overline{\mathbb{G}}(0) \times (1 - s\fund{\tau}/2)}{\left(1 - s/s_{\text{P}+}\right)\left(1 - s/s_{\text{P}-}\right)}
    \odot \begin{bmatrix} 1 - s/s_{\text{T}'\text{q}-} & 1 + s\fund{\tau}/2 \\ 1 + s\fund{\tau}/2 & 1 - s/s_{\text{T}'\text{q}+} \end{bmatrix},
\end{equation}
with
\begin{align}
    \overline{\mathbb{G}}(0) &= \frac{r \mathcal{F}_\Gamma^2}{\pi^2} \begin{bmatrix}
    \cosh{\Gamma} + \cos{2\pump{\psi}} \sinh{\Gamma} -r &
    -\sin{2\pump{\psi}} \sinh{\Gamma} \\
    -\sin{2\pump{\psi}} \sinh{\Gamma} &
    \cosh{\Gamma} - \cos{2\pump{\psi}} \sinh{\Gamma} -r \end{bmatrix} \\
    &\simeq \frac{1/\fund{\tau}}{\kappa^2 - g^2} \begin{bmatrix} \kappa + g\cos{2\pump{\psi}} & -g\sin{2\pump{\psi}} \\ -g\sin{2\pump{\psi}} & \kappa - g\cos{2\pump{\psi}}
    \end{bmatrix}
\end{align}
and
\begin{align}
    s_{\text{T}'\text{q}\mp} &= \frac{2}{\fund{\tau}} \times \frac{\mp(r - \cosh{\Gamma}) + \cos{2\pump{\psi}} \sinh {\Gamma} }{r +\cosh{\Gamma} \pm \cos{2\pump{\psi}} \sinh{\Gamma}} \\
    &\simeq -\kappa \mp g\cos{2\pump{\psi}}.
\end{align}
By comparing with the approximations in \cref{sec:Gsupp}, it is evident that to leading order in the decay and gain rates, $\underline{\symsf{G}}$ and $\overline{\symsf{G}}$ coincide, as do $\underline{\mathbb{G}}$ and $\overline{\mathbb{G}}$.
The leading-order approximation scheme overestimates the dc closed loop gain in the sine quadrature, $\overline{\mathbb{G}}_{22}(0)$, by about \qty{6}{\%} for the amplifier parameters in \cref{tab:vopo} (but with $\pump{\psi} = 0$); as we noted previously, the scheme underestimates the analogous dc internal gain element, $\underline{\mathbb{G}}_{22}(0)$, by about \qty{6}{\%} for the same parameters.

\section{Decompositions of sideband and quadrature transfer functions}
\label{sec:svd}

The sideband transfer functions that we have considered are generically of the form
\begin{equation}
    \symsf{B} = \begin{bmatrix} u\rme^{-\rmi\xi} & v\rme^{-\rmi\phi} \\ v\rme^{+\rmi\phi} & u\rme^{+\rmi\xi} \end{bmatrix}
\end{equation}
with $u$ and $v$ possibly complex, and with $\xi$ and $\phi$ both real.
This can be factorized into
\begin{equation}
    \symsf{B} =
        \begin{bmatrix} \rme^{-\rmi(\xi+\phi)/2} & 0 \\ 0 & \rme^{+\rmi(\xi+\phi)/2} \end{bmatrix}
        \begin{bmatrix} u & v \\ v & u \end{bmatrix}
        \begin{bmatrix} \rme^{-\rmi(\xi-\phi)/2} & 0 \\ 0 & \rme^{+\rmi(\xi-\phi)/2} \end{bmatrix}
\end{equation}
and hence, upon transformation into the quadrature representation $\mathbb{B} = \symsf{A}\?\symsf{B}\?\symsf{A}^\dagger$, one finds
\begin{equation}
    \mathbb{B} =
        \begin{bmatrix*}[r] \cos\tfrac{1}{2}(\xi+\phi) & \sin\tfrac{1}{2}(\xi+\phi) \\ -\sin\tfrac{1}{2}(\xi+\phi) & \cos\tfrac{1}{2}(\xi+\phi) \end{bmatrix*}
        \begin{bmatrix} u + v & 0 \\ 0 & u - v \end{bmatrix}
        \begin{bmatrix*}[r] \cos\tfrac{1}{2}(\xi-\phi) & \sin\tfrac{1}{2}(\xi-\phi) \\ -\sin\tfrac{1}{2}(\xi-\phi) & \cos\tfrac{1}{2}(\xi-\phi) \end{bmatrix*}.
        \label{eq:BqD}
\end{equation}
It is therefore possible to interpret this factorization of $\mathbb{B}$ as a sequence comprising a rotation of the quadratures, a (possibly complex) gain $u \pm v$ in each quadrature, and then a final rotation of the quadratures (see also the discussion in Ref.~\cite[App.~A]{McCuller:2021mbn}).
\end{widetext}

Sometimes it is desirable to arrive at a singular value decomposition on $\mathbb{B}$.
Returning to \cref{eq:BqD}, if $u$ and $v$ are both real and $u \ge |v|$, \cref{eq:BqD} is already such a decomposition.
One example is the parametric amplification process $\xtal{\mathbb{H}} = \symsf{A}\?\xtal{\symsf{H}}\?\symsf{A}^\dagger$ (\cref{eq:HX hyperbolic}), which has singular values $\rme^{+\Gamma}$ and $\rme^{-\Gamma}$.
In cases where the gains $u \pm v$ are complex, which will in general be the case for the quadrature transfer functions $\mathbb{R}$, $\mathbb{T}$, etc., it is straightforward to define $u + v = \Sigma_\text{c} \rme^{\rmi\sigma_\text{c}}$ and $u - v = \Sigma_\text{s} \rme^{\rmi\sigma_\text{s}}$, with $\Sigma_\text{c}$ and $\Sigma_\text{s}$ both nonnegative and $\sigma_\text{c}$ and $\sigma_\text{s}$ both real, whereupon one can factorize $\begin{bmatrix} u + v & 0 \\ 0 & u - v \end{bmatrix}$ into either the left or right multiplication of $\begin{bmatrix} \Sigma_\text{c} & 0 \\ 0 & \Sigma_\text{s} \end{bmatrix}$ by $\begin{bmatrix} \rme^{\rmi\sigma_\text{c}} & 0 \\ 0 & \rme^{\rmi\sigma_\text{s}} \end{bmatrix}$.

\bibliography{opa.bib}

\end{document}